\documentclass{LMCS}

\def\doi{8 (1:13) 2012}
\lmcsheading%
{\doi}
{1--33}
{}
{}
{Apr.~\phantom08, 2010}
{Feb.~27, 2012}
{}

\usepackage{latexsym}
\usepackage[vlined,ruled]{algorithm2e}
\usepackage[bf,small]{caption2}
\usepackage[all]{xy}

\ifx\pdfoutput\undefined
\usepackage{graphicx}
\else
\usepackage{graphicx}
\usepackage{epstopdf}
\fi 
\usepackage{enumerate,hyperref}

\newtheorem{theorem}{Theorem}[section]
\renewcommand{\phi}{\varphi}

\newcommand{\standout}[1]{\noindent \\ \textbf{#1}}
\newcommand\hide[1]{}

\newcounter{mycommentcounter}

\newcommand{\zug}[1]{\langle #1  \rangle}
\newcommand{\rzug}[1]{{\scriptstyle\langle} #1  {\scriptstyle\rangle}}

\newcommand\buchi{B\"uchi }

\newcommand{\natnum}{\mbox{$I\!\!N$}}
\newcommand\deqarrow{\ensuremath{\stackrel{0}{\rightarrow}}}
\newcommand\descarrow{\stackrel{1}{\rightarrow}}
\newcommand\vararrow[1]{\stackrel{#1}{\rightarrow}}
\newcommand{\N}{\mbox{I$\!$N}}

\newcommand{\A}{{\mathcal A}}
\newcommand{\B}{{\mathcal B}}
\newcommand{\G}{{\mathcal G}}
\newcommand{\init}{{1}}

\renewcommand{\graph}{\widetilde}

\newcommand{\arc}{\bar}
\newcommand{\superg}{\widehat}
\newcommand{\supergFD}{\superg{Q}_L}

\newcommand{\Y}[2]{\ensuremath{Y({\graph{#1},\graph{#2}})}}
\newcommand{\Z}[2]{\ensuremath{Z({\superg{#1},\superg{#2}})}}

\newcommand{\abs}[1]{\lvert#1\rvert}

\def\squarebox#1{\hbox to #1{\hfill\vbox to #1{\vfill}}}
\renewcommand{\qed}{\hspace*{\fill}
          \vbox{\hrule\hbox{\vrule\squarebox{.667em}\vrule}\hrule}\smallskip}

\renewenvironment{proof}{\begin{trivlist}
\item[\hspace{\labelsep}{\bf\noindent Proof: }]
}{\qed\end{trivlist}}
\SetKwFunction{LJB}{LJB}
\SetKwFunction{SGS}{SingleGraphSearch}
\SetKwFunction{DGS}{DoubleGraphSearch}

\begin{document}
%\scalefont{0.96}
\title{B\"uchi Complementation and Size-Change Termination\rsuper*}

\author[S.~Fogarty]{Seth Fogarty\rsuper a}
\address{{\lsuper{a,b}}Department of Computer Science, Rice University, Houston, TX}
\email{sfogarty@gmail.com, vardi@cs.rice.edu}
\thanks{{\lsuper a}Work supported in part by NSF
grants CCR-0124077, CCR-0311326, CCF-0613889, ANI-0216467, and CCF-0728882, by
BSF grant 9800096, and by a gift from Intel.}
\author[M.~Y.~Vardi]{Moshe Y.~Vardi\rsuper b}
%\address{\vskip-6 pt}
%\email{vardi@cs.rice.edu}

\keywords{\buchi Complementation, Model Checking, Formal Verification, Automata,
\buchi Automata}
\subjclass{D.2.4}
\titlecomment{{\lsuper*}Earlier version appeared in TACAS09}

\begin{abstract}

We compare tools for complementing nondeterministic \buchi automata with a
recent termination-analysis algorithm. Complementation of \buchi automata is a
key step in program verification. Early constructions using a Ramsey-based
argument have been supplanted by rank-based constructions with exponentially
better bounds.  In 2001 Lee et al.~ presented the size-change termination (SCT)
problem, along with both a reduction to \buchi automata and a Ramsey-based
algorithm. The Ramsey-based algorithm was presented as a more practical
alternative to the automata-theoretic approach, but strongly resembles the
initial complementation constructions for \buchi automata. 

We prove that the SCT algorithm is a specialized realization of the Ramsey-based complementation
construction. To do so, we extend the Ramsey-based complementation construction to provide a
containment-testing algorithm.  Surprisingly, empirical analysis suggests that despite the massive
gap in worst-case complexity, Ramsey-based approaches are superior over the domain of SCT problems.
Upon further analysis we discover an interesting property of the problem space that both explains
this result and provides a chance to improve rank-based tools.  With these improvements, we show
that theoretical gains in efficiency of the rank-based approach are mirrored in empirical
performance.

\end{abstract}
\maketitle
%\pagenumbering{arabic}
%\pagestyle{plain}

\section{Introduction}\label{Sect:Introduction}
The automata-theoretic approach to formal program verification reduces questions
about program adherence to a specification to questions about language
containment. Representing liveness, fairness, or termination properties requires
finite automata that operate on infinite words.  One automaton, $\A$, encodes
the behavior of the program, while another automaton, $\B$, encodes the formal
specification.  To ensure adherence, verify that the intersection of $\A$ with
the complement of $\B$ is empty.  Finite automata on infinite words are
classified by their acceptance condition and transition structure. We consider
here nondeterministic \buchi automata, in which a run is accepting when it
visits at least one accepting state infinitely often. For these automata, the
complementation problem is known to involve an exponential blowup \cite{Mic88}.  Thus the most
difficult step in checking containment is constructing the complementary
automata $\overline{\B}$.  

The first complementation constructions for nondeterministic \buchi automata
employed a Ramsey-based combinatorial argument to partition the set of
all infinite words into a finite set of omega-regular languages. Proposed by \buchi in
1962 \cite{Buc62}, this construction was shown in 1987 by Sistla, Vardi, and
Wolper to be implementable with a blow-up of $2^{O(n^2)}$ \cite{SVW85}. This
brought the complementation problem into singly-exponential blow-up, but left a
gap with the $2^{\Omega(n\log n)}$ lower bound proved by Michel \cite{Mic88}.

The gap was tightened in 1988, when Safra described a $2^{O(n\log n)}$
construction \cite{Saf88}. Work since then has focused on improving the
practicality of $2^{O(n\log n)}$ constructions, either by providing simpler
constructions, further tightening the bound \cite{Sch09}, or improving the derived
algorithms. In 2001, Kupferman and Vardi employed a rank-based analysis of \buchi
automata to simplify complementation \cite{KV01c}.  Recently, Doyen and Raskin
have demonstrated the utility of using a subsumption technique in the
rank-based approach, providing a direct universality checker that scales to
automata several orders of magnitude larger than previous tools \cite{DR09}. 

Separately, in the context of program termination analysis, Lee, Jones, and
Ben-Amram presented the size-change termination (SCT) principle in 2001
\cite{LJB01}.  This principle states that, for domains with well-founded values,
if every infinite computation contains an infinitely decreasing value sequence,
then no infinite computation is possible. Lee et al.~ describe a method of
size-change termination analysis and reduce this problem to the containment of
two \buchi automata. Stating the lack of efficient \buchi containment solvers,
they also propose a Ramsey-based combinatorial solution that captures all
possible call sequences in a finite set of graphs.  The Lee, Jones, and
Ben-Amram (LJB) algorithm was provided as a practical alternative to reducing
the verification problem to \buchi containment, but bears a striking resemblance
to the 1987 Ramsey-based complementation construction \cite{SVW85}.

In this paper we show that the LJB algorithm for deciding SCT
\cite{LJB01} is a specialized implementation of the 1987 Ramsey-based
complementation construction \cite{SVW85}. Section \ref{Sect:Prelim} presents
the background and notation for the paper.  Section \ref{Sect:SCT_vs._Ramsey}
expands the Ramsey-based complementation construction into a containment
algorithm, and then presents the proof that the LJB algorithm is a specialized
realization of this Ramsey-based containment algorithm.  In Section \ref{Sect:Exps}, we
empirically explore Lee et al.'s intuition that Ramsey-based algorithms are more
practical than \buchi complementation tools on SCT problems. Initial
experimentation does suggest that Ramsey-based tools are superior on SCT
problems. This is surprising, as the worst-case complexity of the LJB algorithm
is significantly worse than that of rank-based tools.  Investigating this
discovery in Section \ref{Sect:Rev-Det}, we note that it is natural for SCT problems to be
reverse-deterministic, and that for reverse-deterministic problems the
worst-case bound for Ramsey-based algorithms matches that of the rank-based
approach. This suggests improving the rank-based approach in the face of reverse
determinism. Indeed, we find that reverse-deterministic SCT problems have a
maximum rank of 2, collapsing the complexity of rank-based complementation to
$2^{O(n)}$.  Revisiting our experiments, we discover that with this improvement rank-based tools are
superior on the domain of SCT problems. To further explore the phenomena, we generate a set of
non-reverse-deterministic SCT problems from monotonicity constraint systems, a more complex
termination problem.  We conclude with a discussion in Section
\ref{Sect:Conclusion}. 

\section{Preliminaries} \label{Sect:Prelim} 

In this section we review the relevant details of the \buchi complementation and
size-change termination, introducing along the way the notation used throughout
this paper.  A \emph{nondeterministic \buchi automaton on infinite words} is a
tuple $\B=\zug{\Sigma, Q, Q^{in}, \rho, F}$, where $\Sigma$ is a finite nonempty
alphabet, $Q$ a finite nonempty set of states, $Q^{in} \subseteq Q$ a set of
initial states, $F \subseteq Q$ a set of accepting states, and $\rho~:~Q \times
\Sigma \rightarrow 2^Q$ a nondeterministic transition function. We lift the
$\rho$ function to sets of states and words of arbitrary length as follows.
Given a set of states $R$, define $\rho(R, \sigma)$ to be $\bigcup_{q \in R}
\rho(q,\sigma)$. Inductively, given a set of states $R$: let
$\rho(R,\epsilon)=R$, and for every word $w=\sigma_0...\sigma_n$ let $\rho(R,w)$ be
defined as $\rho(\rho(R,\sigma_0),\sigma_1...\sigma_n)$. 

A {\em run} of a \buchi automaton $\B$ on a word $w \in \Sigma^\omega$ is a
infinite sequence $r=q_0q_1... \in Q^\omega$ such that $q_0 \in
Q^{in}$ and, for every $i \geq 0$, we have $q_{i+1} \in \rho(q_i, w_i)$.  A run
is \emph{accepting} iff $q_i \in F$ for infinitely many $i \in \natnum$.  A word
$w \in \Sigma^\omega$ is accepted by $\B$ if there is an accepting run of $\B$
on $w$.  The words accepted by $\B$ form the language of $\B$, denoted by
$L(\B)$.  Correspondingly, a {\em path} in $\B$ from $q$ to $r$ on a word $w \in
\Sigma^+$ is a finite sequence $r=q_0...q_n \in Q^+$ such that
$q_0=q$, $q_n=r$, and, for every $i \in \{0...n-1\}$, we have $q_{i+1} \in
\rho(q_i, w_i)$.  A path is {\em accepting} if some state in the path is in $F$. 

\begin{figure}
\begin{center}
{\includegraphics[clip=true, trim = 0.1in 1.2in 0.1in 0.1in, angle=90,height=0.25\linewidth]{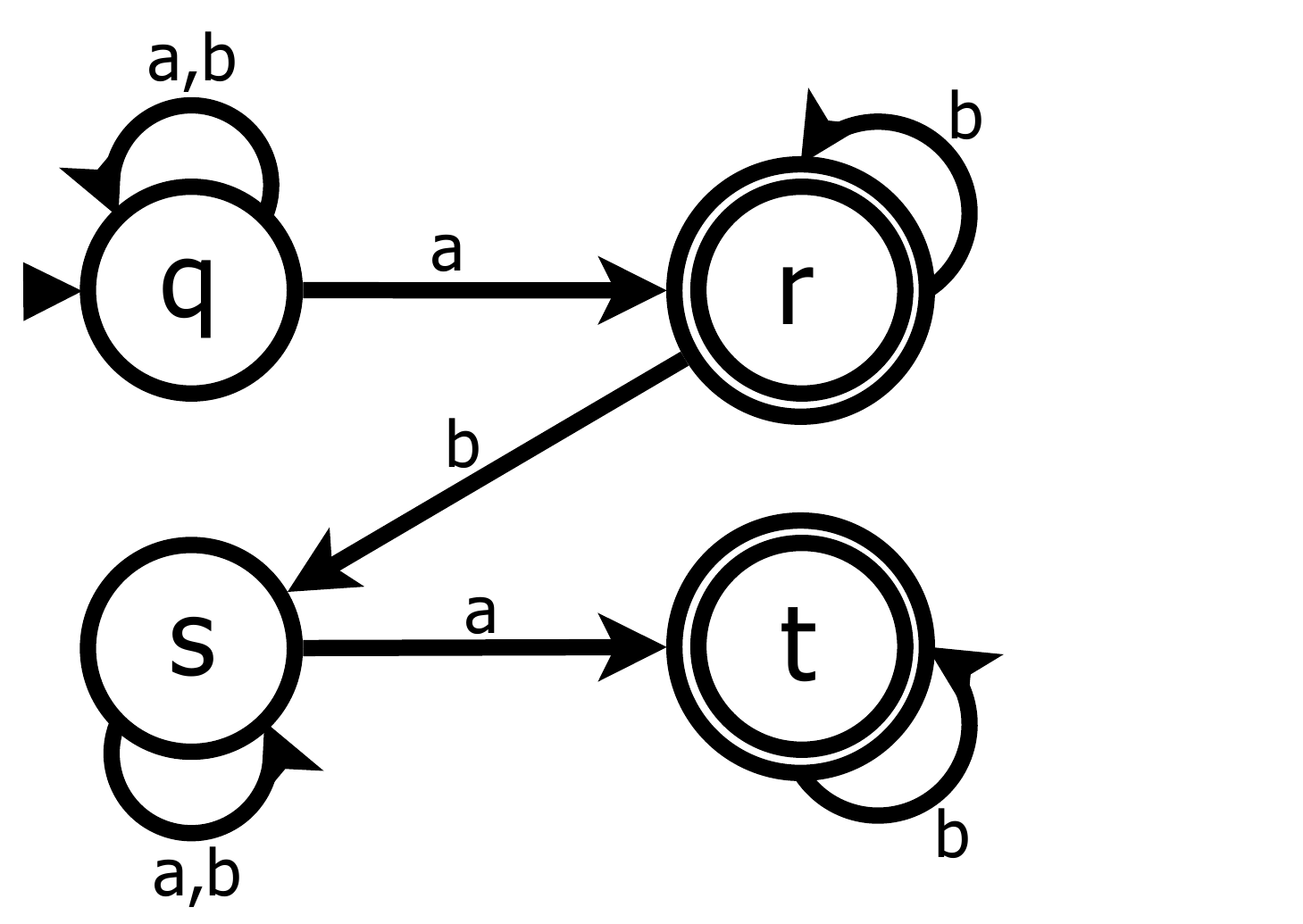}}
\end{center}
\caption{An automaton which accepts words with finitely many $a$'s.}\label{fig:automaton}
\end{figure}

\begin{exa}
An example automaton is shown in Figure \ref{fig:automaton}, which accepts words with a finite, but
non-zero, number of
$a$'s. The automaton waits in $q$ and guesses when it has seen the last $a$, transitioning on that
$a$ to $r$. If the automaton moves to $r$ prematurely, it can transition to $s$ before it encounters
any remaining $a$'s to continue the run. From $s$, it can guess once again when it has seen the last
$a$, transitioning this time to $t$.
\end{exa}

A \buchi automaton $\A$ is contained in a \buchi automaton $\B$ iff $L(\A)
\subseteq L(\B)$, which can be checked by verifying that the intersection of
$\A$ with the complement $\overline{\B}$ of $\B$ is empty: $L(\A) \cap
L(\overline{\B})=\emptyset$.  We know that the language of an automaton is
non-empty iff there are states $q \in Q^{in},~ r \in F$ such that there is a
path from $q$ to $r$ and an accepting path from $r$ to itself.  The initial path
is called the prefix, and the combination of the prefix and cycle is called a
\emph{lasso} \cite{Var07a}.  Furthermore, the intersection of two automata can be
constructed, having a number of states proportional to the product of the number
states of the original automata \cite{Cho74}.
Thus, the most computationally demanding step is constructing the complement of
$\B$.  In the formal verification field, existing empirical work has focused on
the simplest form of containment testing, \emph{universality} testing, where
$\A$ is the universal automaton \cite{DR07,TV05}.

\subsection{Ramsey-Based Universality}\label{Sect:Ramsey-Based_Universality}

When \buchi introduced these automata in 1962, he described a complementation
construction involving a Ramsey-based combinatorial argument \cite{Buc62}. We describe an
optimized implementation presented in 1987 \cite{SVW85}.  To construct the complement of
$\B=\zug{\Sigma, Q, Q^{in}, \rho, F}$, where  $Q=\{q_0,...,q_{n-1}\}$, we
construct a set $\graph{Q}_\B$ whose elements capture the essential behavior of
$\B$.  Each element corresponds to an answer to the following question:

\begin{verse}
Given a finite nonempty word $w$, for every two states $q,r \in Q$:
\begin{enumerate}
\item Is there a path in $\B$ from $q$ to $r$ over $w$?
\item If so, is some such path accepting?
\end{enumerate}
\end{verse}

Define $Q'=Q\times\{0,1\}\times Q$, and $\graph{Q}_\B$ to be the subset of $2^{Q'}$
whose elements, for every $q, r \in Q$, do not contain both $\zug{q,0,r}$ and $\zug{q,1,r}$.  Each
element of $\graph{Q}_\B$ is a $\{0,1\}$-arc-labeled graph on $Q$. An arc represents a path in $\B$,
and the label is $1$ if the path is accepting. Note that there are $3^{n^2}$ such graphs.  With each
graph $\graph{g} \in \graph{Q}_\B$ we associate a language $L(\graph{g})$, the set of words for
which the answer to the posed question is the graph encoded by $\graph{g}$. 

\begin{defi}\label{X_Describes}
Let $\graph{g} \in \graph{Q}_\B$ and $w \in \Sigma^+$. Then $w \in L(\graph{g})$
iff, for all pairs of states $q, r \in Q$:
\begin{enumerate}[(1)]
\item $\zug{q,a,r} \in \graph{g},~ a \in \{0, 1\}$, iff there is a path in $\B$
from $q$ to $r$ over $w$.
\item $\zug{q,1,r} \in \graph{g}$ iff there is an accepting path in $\B$ from
$q$ to $r$ over $w$.
\end{enumerate} 
\end{defi}

\begin{exa}
Three graphs from $\graph{Q}_\B$ are shown in Figure \ref{fig:automaton}. All graphs have a
non-empty language. The word $a$ is in the language of the first graph, the word $b$ is in the
language of the second graph, and the word $ab$ is in the language of the third graph. 
\end{exa}

\begin{figure}[tb]
\begin{center}
{\includegraphics[clip=true, trim = 1.1in 6.4in 1.1in 1.1in, height=0.3\linewidth]{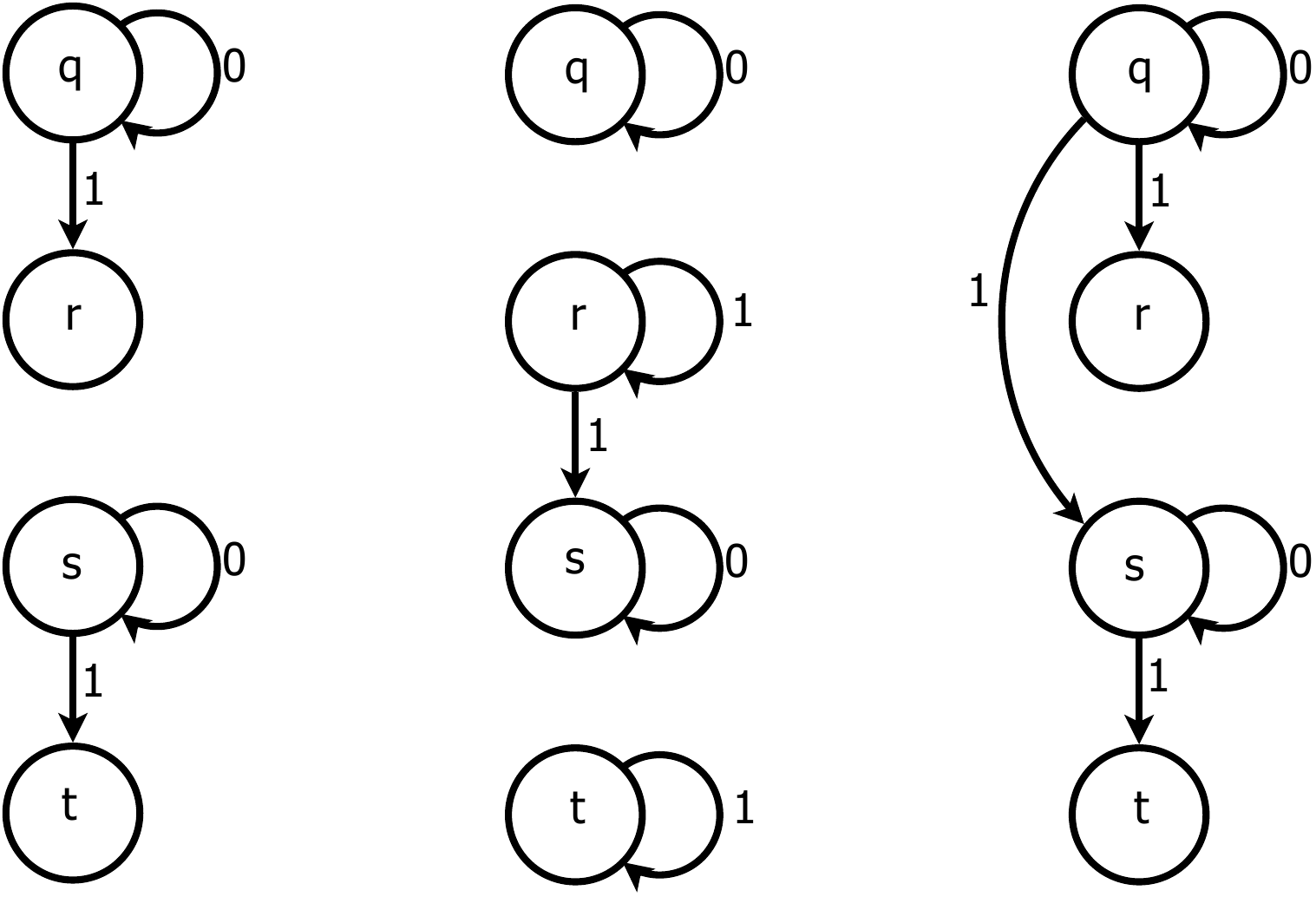}}
\end{center}
\caption{Three graphs in $\graph{Q}_\B$ for the automaton of Figure \ref{fig:automaton}. From left to
right, the graph describing the word $a$, the graph describing the word $b$, and the graph
describing the word $ab$.}  \label{fig:svw-graph}
\end{figure}

\begin{lem}\label{X_Partitions}{\rm \cite{Buc62,SVW85}}\  
\begin{enumerate}[\em(1)]
\item $\{L(\graph{g})\mid \graph{g} \in \graph{Q}_\B\}$ is a partition of
$\Sigma^+$
\item If $u \in L(\graph{g}),~ v \in L(\graph{h}),\text{ and }uv\in
L(\graph{k}), \text{ then } L(\graph{g}) \cdot L(\graph{h}) \subseteq
L(\graph{k})$
\end{enumerate}
\end{lem}

The languages $L(\graph{g})$, for the graphs $\graph{g} \in \graph{Q}_\B$, form
a partition of $\Sigma^+$. With this partition of $\Sigma^+$ we can devise a
finite family of $\omega$-languages that cover $\Sigma^\omega$. For every
$\graph{g},~ \graph{h} \in \graph{Q}_\B$, let $\Y{g}{h}$~ be the
$\omega$-language $L(\graph{g})\cdot L(\graph{h})^\omega$. Say that a language
$\Y{g}{h}$ is \emph{proper} if $\Y{g}{h}$ is non-empty, $L(\graph{g})\cdot
L(\graph{h}) \subseteq L(\graph{g})$, and $L(\graph{h})\cdot L(\graph{h})
\subseteq L(\graph{h})$.  There are a finite, if exponential, number of such
languages. A Ramsey-based argument shows that every infinite string belongs to
a language of this form, and that $\overline{L(\B)}$ can be expressed as the
union of languages of this form. 

\begin{lem}\label{Proper_Covers}{\rm \cite{Buc62,SVW85}}\ 
$\Sigma^\omega = \bigcup\{\Y{g}{h}~|~\Y{g}{h}\text{ is proper}\}$
\end{lem}
\begin{proof} 
The proof is based on Ramsey's Theorem.  Consider an infinite word
$w=\sigma_0\sigma_1...$ By Lemma \ref{X_Partitions}, every prefix of the word
$w$ is in the language of a unique graph $\graph{g}_i$.  Let $k=3^{n^2}$ be the number of graphs.
Thus $w$ defines a partition of $\N$ into $k$ sets $D_1,...,D_k$ such that $i\in
D_l$ iff $\sigma_0...\sigma_{i-1} \in L(\graph{g}_l)$.  Clearly there is some
$m$ such that $D_m$ is infinite.

Similarly, by Lemma \ref{X_Partitions} we can use the word $w$ to define a
partition of all \emph{pairs} of elements $(i,j)$ from $D_m$, where $i < j$. This partition consists
of $k$ sets $C_1,...C_k$, such that $\zug{i,j} \in C_l$ iff $\sigma_i...\sigma_{j-1} \in
L(\graph{g}_l)$. Ramsey's Theorem tells us that, given such a partition, there
exists an infinite subset $\{i_1,i_2,...\}$ of $D_m$ and a $C_n$ such that
 for all pairs of distinct elements $i_j,i_k$, it holds that $\zug{i_j,i_k}\in C_n$. 

This implies that the word $w$ can be partitioned into $$w_1=\sigma_0...\sigma_{i_1-1},~ ~
~ ~ w_2=\sigma_{i_1}...\sigma_{i_2-1},~ ~ ~ ~ w_3=\sigma_{i_2}...\sigma_{i_3-1},~ ~ ~ ~ ...,$$ where
$w_1 \in L(\graph{g}_m)$ and $w_i \in L(\graph{g}_n)$ for $i>1$. By construction,
$\sigma_0...\sigma_{i_j-1} \in L(\graph{g}_m)$ for every $i_j$, and thus we have that $w_1w_2 \in
L(\graph{g}_m)$. In addition, as $\sigma_{i_j}...\sigma_{i_k-1} \in L(\graph{g}_n)$ for every pair
$i_j, i_k$, we have that $w_2w_3 \in L(\graph{g}_n)$.  By Lemma
\ref{X_Partitions}, it follows that $L(\graph{g}_m) \cdot L(\graph{g}_n) \subseteq
L(\graph{g}_m)$, and that $L(\graph{g}_n) \cdot L(\graph{g}_n) \subseteq
L(\graph{g}_n)$, and thus $\Y{g_m}{g_n}$ is proper.  
\end{proof}

Furthermore, each proper language is entirely contained or
entirely disjoint from $L(\B)$. This provides a way to construct the complement
of $L(\B)$: take the union every proper language that is disjoint from $L(\B)$.

\begin{lem}\label{Proper_Disjoint}{\rm \cite{Buc62,SVW85}}\ 
\begin{enumerate}[\em(1)]
\item For $\graph{g}, \graph{h} \in \graph{Q}_\B$, either $\Y{g}{h} \cap L(\B) =
\emptyset$ or $ \Y{g}{h} \subseteq L(\B)$.
\item $\overline{L(\B)}= \bigcup\{\Y{g}{h}~|~\Y{g}{h}$ is proper and $\Y{g}{h} \cap
L(\B) = \emptyset\}$
\end{enumerate} 
\end{lem} 

To obtain the complementary \buchi automaton $\overline{\B}$,  Sistla et al.~
construct, for each $\graph{g}\in\graph{Q}_\B$, a deterministic automata on
finite words, $\B_g$, that accepts exactly $L(\graph{g})$. Using the automata $\B_g$, one can then
construct the complementary automaton $\overline{\B}$ \cite{SVW85}. We can then
use a lasso-finding algorithm on $\overline{\B}$ to prove the emptiness of
$\overline{\B}$, and thus the universality of $\B$.  However, we can avoid an
explicit lasso search by employing the rich structure of the graphs in
$\graph{Q}_\B$. For every two graphs $\graph{g}, \graph{h} \in \graph{Q}_\B$,
determine if $\Y{g}{h}$ is proper.  If $\Y{g}{h}$ is proper, test if it is
contained in $L(\B)$ by looking for a lasso with a prefix in $\graph{g}$ and a
cycle in $\graph{h}$.  $\B$ is universal if every proper $\Y{g}{h}$ is so
contained.

\begin{lem}\label{SVW_Checks}{\rm \cite{SVW85}}
Given an \buchi automaton $\B$ and the set of graphs $\graph{Q}_\B$,
\begin{enumerate}[\em(1)]
\item $\B$ is universal iff for every proper $\Y{g}{h}$, it holds that $\Y{g}{h} \subseteq L(\B)$.
\item Let $\graph{g}, \graph{h} \in \graph{Q}_\B$ be two graphs where $\Y{g}{h}$
is proper. $\Y{g}{h} \subseteq L(\B)$ iff there exists $q \in Q^{in},~ r \in Q,~ a
\in \{0,1\}$ where $\zug{q,a,r} \in \graph{g} $ and $\zug{r,1,r} \in
\graph{h}$.  
\end{enumerate}
\end{lem}

Lemma \ref{SVW_Checks} yields a PSPACE algorithm to determine universality
\cite{SVW85}.  Simply check each $\graph{g}, \graph{h} \in \graph{Q}_\B$.  If
$\Y{g}{h}$ is both proper and not contained in $L(\B)$, then the pair
$(\graph{g}, \graph{h})$ provide a counterexample to the universality of
$\B$. If no such pair exists, the automaton must be universal.

\subsection{Rank-Based Complementation}\label{Rank-Based}

While our focus is mainly on the Ramsey-based approach, in Section \ref{Sect:Rev-Det} we 
look at the rank-based construction described here. If a \buchi automaton $\B$ does not accept a
word $w$, then every run of $\B$ on $w$ must eventually cease visiting accepting states.  The
rank-based construction, foreshadowed in \cite{Kla90} and first introduced in \cite{KV97b}, uses a notion of ranks to track the
progress of each possible run towards this point.  Consider a \buchi automaton $\B=\zug{\Sigma, Q,
Q_\in, \rho, Q_f}$ and an infinite word $w = \omega_0\omega_1...$. The runs of $\B$ on $w$ can be
arranged in an infinite DAG (directed acyclic graph), $G_w=\zug{V,E}$, where 

\begin{iteMize}{$\bullet$}
\item $V \subseteq Q \times \N$ is such that $\zug{q,l} \in V$ iff some run $r$ of
$\B$ on $w$ has $r(l)=q$.
\item $E \subseteq \bigcup_{l \geq 0} (Q \times \{l\}) \times (Q \times
\{l+1\})$ is
$E(\zug{q,l},\zug{q',l+1})$ iff $\zug{q,l} \in V$ and $q' \in \rho(q,
\omega_l)$.
\end{iteMize}

$G_w$, called the \emph{run DAG} of $\B$ on $w$, exactly embodies all possible
runs of $\B$ on $w$. We define a run DAG $G_w$ to be {\em accepting} when there exists
a path in $G_w$ with infinitely many states in $F$. This path corresponds to an
accepting run of $\B$ on $w$. When $G_w$ is not accepting, we say it is a
\emph{rejecting} run DAG.  Say that a node $\zug{q,i}$ of a graph is \emph{finite} if
it has only finitely many descendants, that it is {\em accepting} if $q \in Q_f$, and
that it is \emph{$F$-free} if it is not accepting and does not have 
accepting descendants.  

Given a run DAG $G_w$, we inductively define a sequence of subgraphs by
eliminating nodes that cannot be part of accepting runs. A node that is finite
can clearly not be part of an infinite run, much less an accepting infinite
run. Similarly, a node that is $F$-free may be part of an infinite run, but this
infinite run can not visit an infinite number accepting states.

\begin{iteMize}{$\bullet$}
\item $G_w(0)=G_w$
\item $G_w(2i+1)=G_w(2i) \setminus \{\zug{q,l}~|~\zug{q,l}\text { is finite in }G_w(2i)\}$
\item $G_w(2i+2)=G_w(2i+1) \setminus \{\zug{q,l}~|~\zug{q,l}\text { is }F\text{-free in }G_w(2i)\}$
\end{iteMize}
If the final graph a node appears in is $G_w(i)$, we say that
node is of rank $i$. If the rank of every node is at most $i$, we say $G_w$ has a
rank of $i$.  $\B$ has a maximum rank of $i$ when  every rejecting run DAG of
$\B$ has a maximum rank of $i$.  Note that nodes with odd ranks are removed
because they are $F$-free. Therefore no accepting state can have an odd rank.
Kupferman and Vardi prove that the maximum rank of a rejecting run DAG for every
automaton is bounded by $2\abs{Q}-2$ \cite{KV97b}. This allows us to create an automaton that
guesses the ranking of rejecting run DAG of $\B$ as it proceeds along the
word.

A \emph{level ranking} for an automaton $\B$ with $n$ states is a function $f : Q \rightarrow
\{0...2n-2,\bot\}$, such that if $q \in F$ then $f(q)$ is even or $\bot$. Let $a$ be a letter in
$\Sigma$ and $f, f'$ be two level rankings $f$.  Say that \emph{$f$ covers $f'$ under $a$} when for
all $q$ and every $q' \in \rho(q,a)$, if $f(q) \neq \bot$ then $f'(q') \neq \bot$ and $f'(q') \leq
f(q)$; i.e. no transition between $f$ and $f'$ on $a$ increases in rank. Let $F_r$ be the set of all
level rankings.

\begin{defi}\label{KVDef}
If $\B = \zug{\Sigma, Q, Q^{in}, \rho, F}$ is a \buchi automaton, define
$KV(\B)$ to be the automaton $\zug{\Sigma, F_r \times 2^Q, \rzug{f_{in},\emptyset},
\rho', F_r \times \{\emptyset\}}$, where
\begin{iteMize}{$\bullet$}
\item $f_{in}(q)=2n-2$ for each $q \in Q^{in}$,~ $\bot$ otherwise.
\item Define $\rho' : \zug{F_r \times 2^Q} \times \sigma \rightarrow 2^{\zug{F_r \times 2^Q}}$ to be
\begin{iteMize}{$-$}
\item If $o\neq\emptyset$ then $\rho'(\zug{f,o},\sigma)=\\
\{\rzug{f',o'\setminus d}~|~ f$ covers $f'$ under $\sigma$, 
$o'=\rho(o,\sigma)$, 
$d=\{q~|~f'(q)\text{ odd}\}\}$.
\item If $o=\emptyset$ then $\rho'(\rzug{f,o},\sigma)=\\
\{\rzug{f',o'}~|~ f$ covers $f'$ under $a$, 
$o' = \{q~|~f'(q)\text{ even}\}\}$.
\end{iteMize}
\end{iteMize}
\end{defi}

\begin{lem}\label{KV_Complement}{\rm \cite{KV01}}
For every \buchi automaton $\B$, $L(KV(\B))=\overline{L(\B)}$.
\end{lem}

This automaton tracks the progress of $\B$ along a word $w=\sigma_0\sigma_1...$ by
attempting to find an infinite series $f_0f_1...$ of level rankings. We start
with the most general possible level ranking, and ensure that every rank $f_i$
covers $f_{i+1}$ under $\sigma_i$. Every run has a non-increasing rank, and so must
eventually become trapped in some rank. To accept a word, the automaton requires
that each run visit an odd rank infinitely often. Recall that accepting states
cannot be assigned an odd rank. Thus, for a word rejected by $\B$, every run can
eventually become trapped in an odd rank.  Conversely, if there is an accepting
run that visits an accepting node infinitely often, that run cannot visit an odd
rank infinitely often and the complementary automaton rejects it.

An algorithm seeking to refute the universality of $\B$ can look for a lasso in the state-space of
the rank-based complement of $\B$. A classical approach is Emerson-Lei backward-traversal nested
fixpoint $\nu Y. \mu X . (Pre(X) \cup ( Pre(Y) \cap F ))$ \cite{EL86}.  This nested fixpoint employs
the observation that a state in a lasso can reach an arbitrary number of accepting states. The outer
fixpoint iteratively computes sets $Y_0, Y_1,...$ such that $Y_i$ contains all states with a path
visiting $i$ accepting states.  Universality is checked by testing if $Y_\infty$, the set of all
states with a path visiting arbitrarily many accepting states, intersects $Q^{in}$.  The strongest
algorithm implementing this approach, from Doyen and Raskin, takes advantage of the presence of a
subsumption relation in the rank-based construction: one state $\rzug{f, o}$ subsumes another
$\rzug{f', o'}$ iff: $f'(x) \leq f(x)$ for every $x \in Q$; $o' \subseteq o$; and $o = \emptyset$
iff $o' = \emptyset$.  When computing sets in the Emerson-Lei approach, it is sufficient to store
only the maximal elements under this relation. Furthermore, the predecessor operation for a single
state and letter results in at most two incomparable elements. This algorithm has scaled to automata
an order of magnitude larger than other approaches \cite{DR07}.

\subsection{Size-Change Termination}\label{SCT}

In \cite{LJB01} Lee et al.~ proposed the size-change termination (SCT)
principle for programs: ``If every infinite computation would give rise to an infinitely
decreasing value sequence, then no infinite computation is possible.'' The
original presentation concerned a first-order pure functional language,
where every infinite computation arises from an infinite call sequence and
values are always passed through a sequence of parameters.

Proving that a program is size-change terminating is done in two phases. The
first extracts from a program a set of size-change graphs, $\G$, containing
guarantees about the relative size of values at each function call site. The
second phase, and the phase we focus on, analyzes these graphs to determine if
every infinite call sequence has a value that descends infinitely along a
well-ordered set. For an excellent discussion of the abstraction of functional
language semantics, refer to \cite{JB04}.  We consider here a set $H$ of
functions, and denote the parameters of a function $f$ by $P(f)$.

\begin{defi}
A \emph{size-change graph} (SCG) from function $f_1$ to function $f_2$, written
$G : f_1 \rightarrow f_2$, is a bipartite $\{0,1\}$-arc-labeled graph from
the parameters of $f_1$ to the parameters of  $f_2$, where $G \subseteq P(f_1)
\times \{0,1\} \times P(f_2)$ does not contain both $x \descarrow y$ and $x
\deqarrow y$.
\end{defi}

\begin{figure}[tb]
\begin{center}
{\includegraphics[clip=true, trim = 1.1in 9.4in 3.7in 1.0in, height=0.16\linewidth]{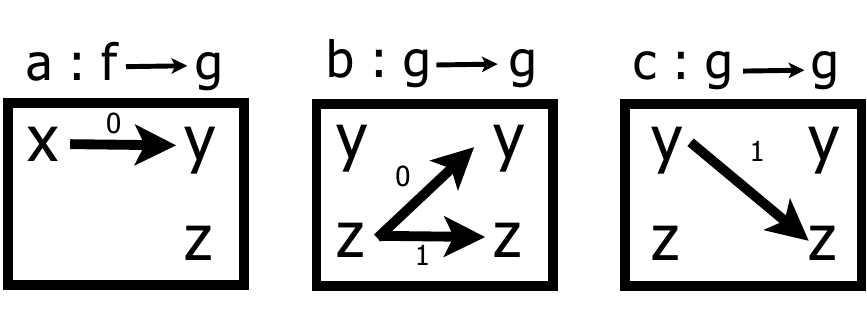}}
\end{center}
\caption{Size-Change Graphs: A size-change problem with two functions, $f$ and $g$, and three call
sites: a call $a$ to $g$ occurring in the body of $f$, and two recursive calls, $b$ and $c$, from $g$ to itself.}\label{fig:scg}
\end{figure}

Size-change graphs capture information about a function call. An arc $x
\descarrow y$ indicates that the value of $x$ in the function $f_1$ is strictly
greater than the value passed as $y$ to function $f_2$.  An arc $x \deqarrow y$
indicates that $x$'s value is greater than or equal to the value given to $y$.
We assume that all call sites in a program are reachable from the entry points of the
program\footnote{The implementation provided by Lee et al.
\cite{LJB01} also make this assumption, and in the presence of unreachable
functions size-change termination may be undetectable.}.

A \emph{size-change termination} (SCT) problem is a tuple $L=\zug{H, P, C, \G}$, where $H$ is a set
of functions, $P$ a mapping from each function to its parameters, $C$ a set of call sites between
these functions, and $\G$ a set of SCGs for $C$.  A call site is written $c : f_1 \rightarrow f_2$
for a call to function $f_2$ occurring in the body of $f_1$.  The size-change graph for a call site
$c : f_1 \rightarrow f_2$ is written as $G_c$. Given a SCT problem $L$, a \emph{call sequence} in
$L$ is a infinite sequence $cs = c_0,c_1,\ldots \in C^\omega$, such that there exists a sequence of
functions $f_0,f_1,\ldots$ where $c_0 : f_0 \rightarrow f_1$, $c_1 : f_1 \rightarrow f_2\ldots$.  A
\emph{thread} in a call sequence $c_0,c_1,\ldots$ is a connected sequence of arcs, $x \vararrow{a}
y, y \vararrow{b} z,\ldots$, beginning in some call $c_i$ such that $x \vararrow{a} y \in G_{c_i}, y
\vararrow{b} z \in G_{c_{i+1}}, \ldots$.  We say that $L$ is \emph{size-change terminating} if every
call sequence contains a thread with infinitely many $1$-labeled arcs.  Note that a thread need not
begin at the start of a call sequence. A sequence must terminate if a well-founded value decreases
infinitely often, regardless of when this decrease begins. Therefore threads can begin in arbitrary
function calls, in arbitrary parameters. We call this the \emph{late-start property} of SCT
problems We revisit this property in Section \ref{Sect:Suffix_Closed}.

\begin{figure}[t]
\begin{center}
{\includegraphics[clip=true, trim = 5.4in 1.0in 1.0in 1.0in, angle = 90, height=0.16\textwidth]{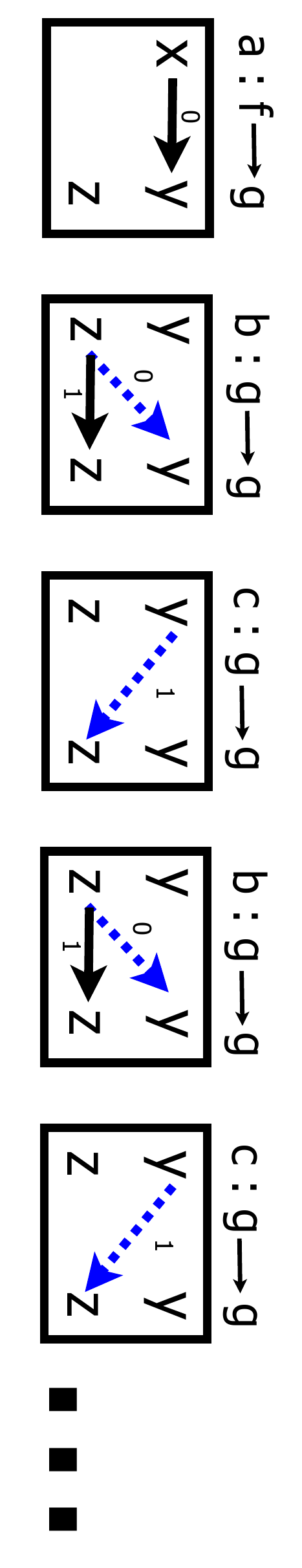}}
\end{center}
\caption{The dotted line forms a prefix of a late-start thread in the call sequence $abcbcbc\cdots$.}\label{fig:late-start}
\end{figure}

\begin{exa}
Three size-change graphs, which will provide a running example for this paper, are presented in
Figure \ref{fig:scg}. The represented problem is size-change terminating. The call sequence
$abcbc\ldots$ is displayed in Figure \ref{fig:late-start}, where a thread of infinite descent
exists, starting in the second graph. This late-start thread proves the sequence terminating.
\end{exa}

Every call sequence can be represented as a word in $C^\omega$, and a SCT
problem reduced to the containment of two $\omega$-languages. The first language
$Flow(L) = \{cs \in C^\omega~|~cs\text{ is a call sequence}\}$, contains all
call sequences. The second language, $Desc(L) =$ $\{cs \in Flow(L)~|~$some
thread in $cs$ has infinitely many 1-labeled arcs$\}$, contains only call
sequences that guarantee termination. A SCT problem $L$ is size-change
terminating if and only if $Flow(L) \subseteq Desc(L)$.

\begin{figure}[tb]
\begin{center}
\begin{tabular}{cc}
\raisebox{0.3in}{{
{\includegraphics[clip=true, trim = 1.0in 1.1in 1.1in 1.0in,
height=0.17\linewidth]{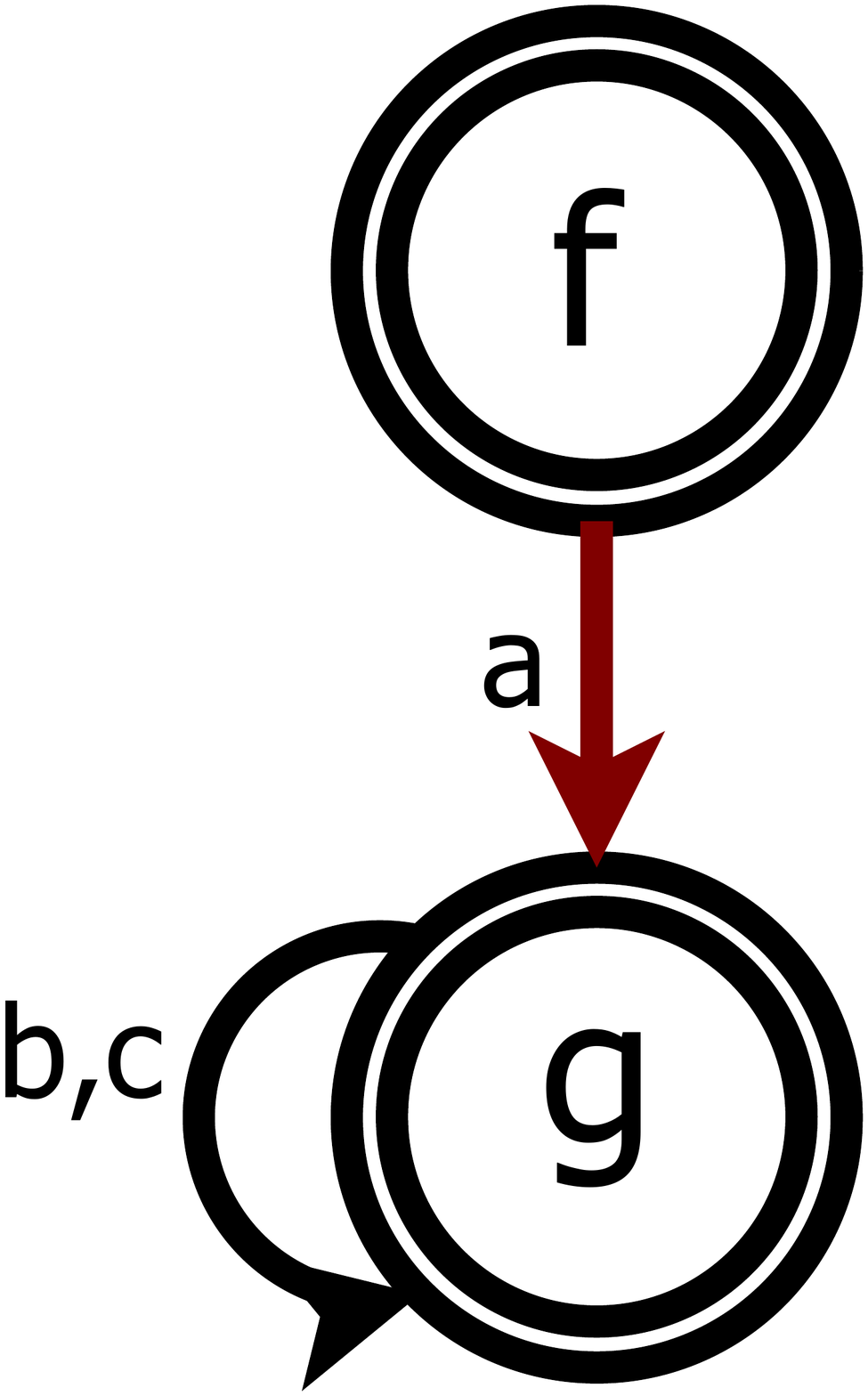}}
}}
\hspace{0.25in}
&
\hspace{0.25in}
{
{\includegraphics[clip=true, trim = 1.0in 3.6in 1.0in 1.0in,
height=0.4\linewidth]{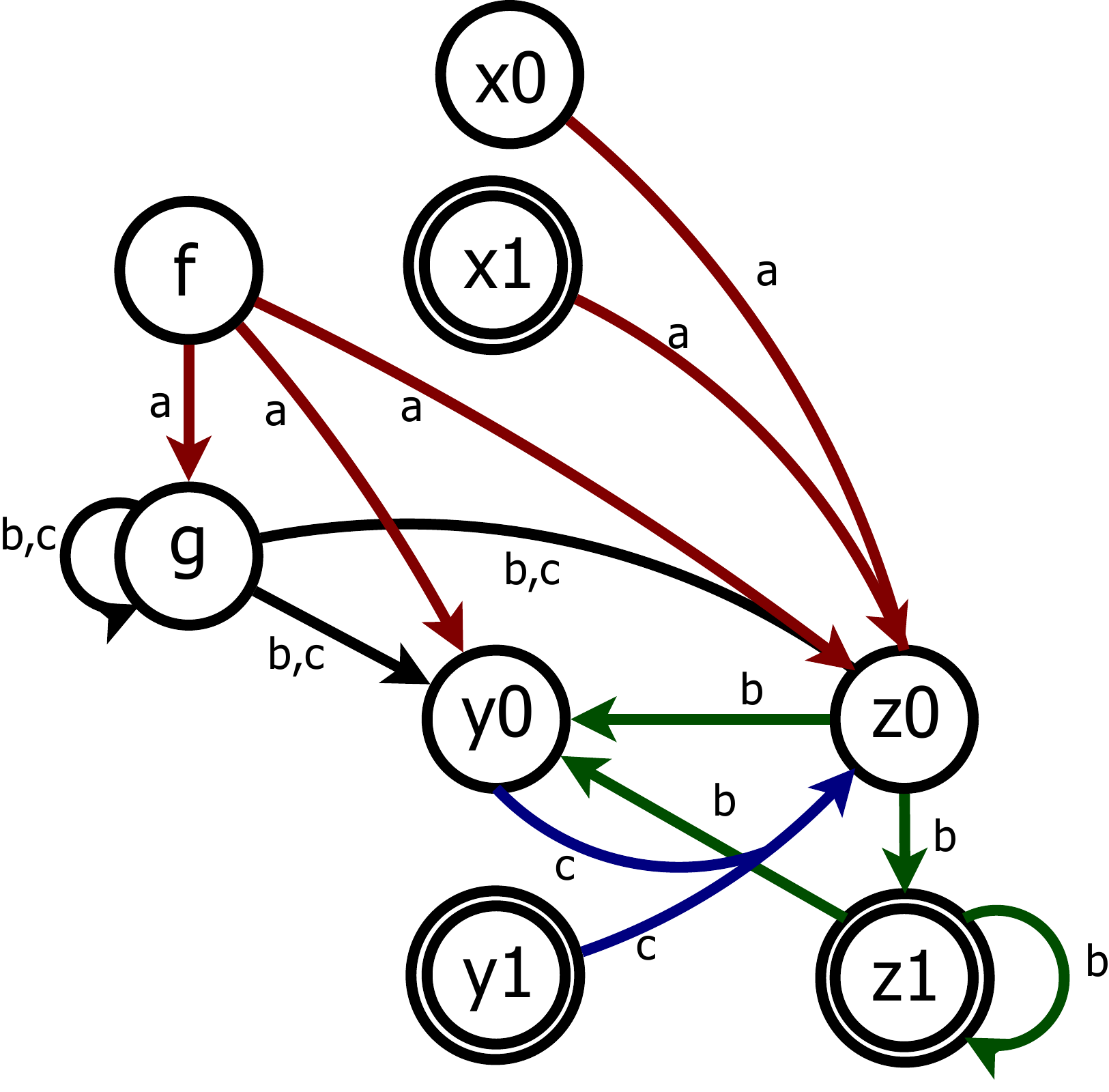}}
}
\end{tabular}
\end{center}
\caption{$\A_{Flow(L)}$ and $\A_{Desc(L)}$: the automata resulting from applying Definition
\ref{LJB_Reduction} to the SCT problem of Figure \ref{fig:scg}.}
\label{fig:ljb-desc}
\end{figure}

Lee et al.~ \cite{LJB01} describe two \buchi automata, $\A_{Flow(L)}$ and
$\A_{Desc(L)}$, that accept these languages. $\A_{Flow(L)}$ is simply the call
graph of the program. $\A_{Desc(L)}$ waits in a copy of the call graph and
nondeterministically chooses the beginning point of a descending thread.  From
there it ensures that a $1$-labeled arc is taken infinitely often. To do so, it
keeps two copies of each parameter, and transitions to the accepting copy only
on a $1$-labeled arc.  Lee et al.~ prove that $L(\A_{Flow(L)})=Flow(L)$, and
$L(\A_{Desc(L)})=Desc(L)$. The automata for our running example are provided in Figure
\ref{fig:ljb-desc}.

\begin{defi}\label{LJB_Reduction} 
$\!\!$\footnote{The original LJB construction \cite{LJB01} restricted starting
states in $\A_{Desc(L)}$to functions.  This was changed to simplify Section
\ref{Sect:SCT_Is_Ramsey}. The modification does not change the accepted
language.}\\
$\A_{Flow(L)} = \zug{C, H, H, \rho_F, H} \textnormal{,~ ~ where}$
\begin{iteMize}{$\bullet$}
\item $\rho_F(f_1, c)=\{f_2~|~c:f_1 \rightarrow f_2\}$
\end{iteMize}
$\A_{Desc(L)} = \zug{C, Q_p\cup H,Q_p \cup H,\rho_D,F} \textnormal{,~ ~ where }$
\begin{iteMize}{$\bullet$}
\item $Q_p=\{\zug{x,r}~|~f\in H,~ x \in P(f),~ r \in \{1, 0\}\},$
\item $\rho_D(f_1, c)=\{f_2~|~c:f_1 \rightarrow f_2\} ~ \cup\, $ $\{\zug{x,0}~|~c:f_1 \rightarrow f_2,~ x\in P(f_2)\}\,$
\item $\rho_D(\zug{x,r}, c)=\{\zug{x',r'}~|~x\vararrow{r'}x' \in \G_c\},$
\item $F=\{\zug{x,1}~|~f \in H,~ x \in P(f)\} $
\end{iteMize}
\end{defi} 

Using the complementation constructions of either Section
\ref{Sect:Ramsey-Based_Universality} or \ref{Rank-Based} and a lasso-finding
algorithm,  we can determine the containment of $\A_{Flow(L)}$ in
$\A_{Desc(L)}$. Lee et al.~ propose an alternative graph-theoretic algorithm,
employing SCGs to encode descent information about entire call sequences.  A
notion of composition is used, where a call sequence $c_0...c_{n-1}$ has a
thread from $x$ to $y$ if and only if the composition of the SCGs for each call,
$G_{c_0};...;G_{c_{n-1}}$, contains the arc $x \vararrow{a} y$. The closure of
$\G$ under the composition operation, called $S$,  is then searched for a
counterexample describing an infinite call sequence with no infinitely
descending thread.

\begin{defi} 
Let $G : f_1 \rightarrow f_2$ and $G' : f_2 \rightarrow f_3$ be two SCGs.
Their composition $G;G'$ is defined as $G'' : f_1 \rightarrow f_3$ where:
\begin{eqnarray*}
G'' &=& \{x \descarrow z~|~x \vararrow{a} y \in G,~ y \vararrow{b} z \in G',\,y \in P(f_2),\, a=1\text{ or }b=1\}\\
&\cup&  \{x \deqarrow z~|~x \deqarrow y \in G,~ y \deqarrow z \in G',~ y\in P(f_2),\text{ and}\\
&&\text{   for all } y',a,b \text{ if }x \vararrow{a} y' \in G \text{ and } y' \vararrow{b} z \in G' \text{ then }a=b=0\}
\end{eqnarray*}
\end{defi}

\begin{figure}[tb]
\begin{center}
{\includegraphics[clip=true, angle = 90, trim = 4.4in 1.1in 1.2in 1.1in,
height=0.16\linewidth]{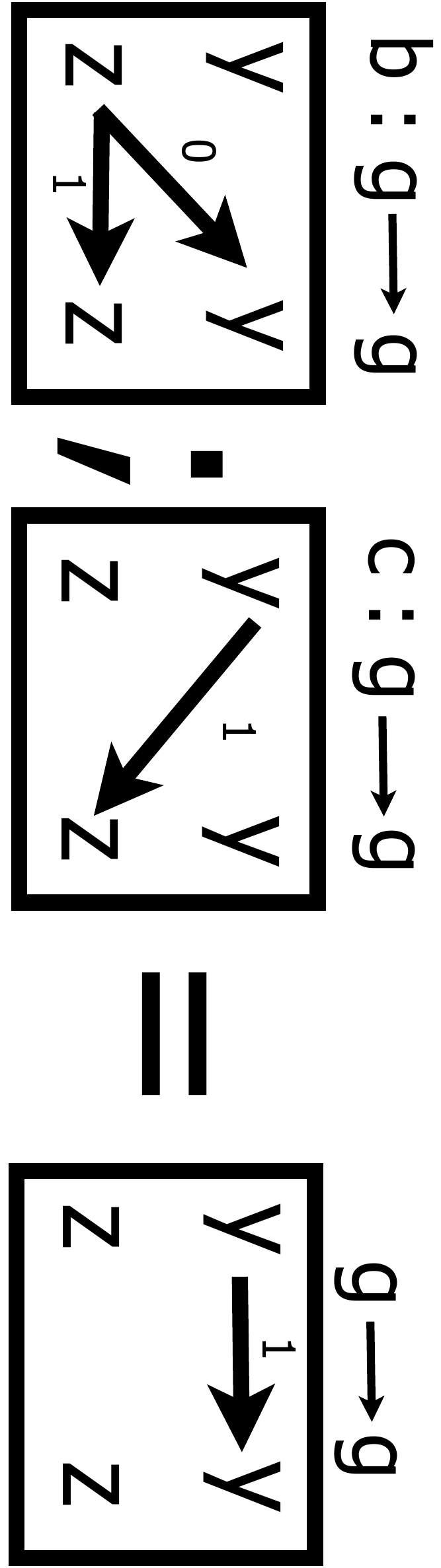}}
\end{center}
\caption{The composition of the SCGs for $b$ and $c$, from Figure \ref{fig:scg}. The
resulting size-change graph is idempotent, contains the arc $y \descarrow y$, and
describes the call sequence of Figure \ref{fig:late-start}.}
\label{fig:scg-comp}
\end{figure}

Using composition, we can focus on a subset of graphs. Say that a graph $G : f \rightarrow f$ is
\emph{idempotent} when $G=G;G$. Each idempotent graph describes a cycle in the call graph, and a
Ramsey-based argument shows that each cycle in the call graph can be accounted for by at least one
idempotent graph. The composition of two graphs is shown in Figure \ref{fig:scg-comp}, which
describes the call sequence in Figure \ref{fig:late-start}.

Algorithm \LJB searches for a counterexample to size-change termination.  First, it iteratively
build the closure set $S$: initialize $S$ as $\G$; and for every $G : f_1 \rightarrow f_2$ and $G' :
f_2 \rightarrow f_3$ in $S$, include the composition $G;G'$ in $S$.  Second, the algorithm  check
every $G : f_1 \rightarrow f_1 \in S$ to ensure that if $G$ is idempotent, then $G$ has an
associated thread with infinitely many 1-labeled arcs. This thread is represented by an arc of the
form $x \descarrow x$. There are pathological SCT problems for which the complexity of Algorithm
\LJB is $2^{O((n/2)^2)}$.

\begin{algorithm}[htb]\label{Alg:LJB}
\caption{\FuncSty{LJB($\zug{H,P,C,\G}$)}}
\DontPrintSemicolon
\KwData{A size-change termination problem $\zug{H,P,C,\G}$.}
\KwResult{Whether or not the problem is size-change terminating.}
\SetKwData{Sv}{S}
  Initialize $\Sv \Leftarrow \G$\;
  \Repeat{$\Sv$ reaches closure}
  {
  \For{\emph{\KwSty{all}} pairs $G : f \rightarrow g$,~~$G' : g \rightarrow h$ in  \Sv} 
  {
    $G'' : f \rightarrow h \Leftarrow G;G'$\;
    Add $G''$ to $\Sv$\;
    \If{$f = h$ and $G'';G''=G''$}
    {
       \If{there does not exist an arc of the form $x \descarrow x$ in $G''$}
       {
			    \Return{Not Terminating}
       }
    }
  }
  }
	\Return{Terminating}
\end{algorithm}

The next theorem, whose proof uses a Ramsey-based argument,
demonstrates the correctness of Algorithm \LJB in determining the size-change
termination of an SCT problem $L=\zug{H,P,C,\G}$.

\begin{theorem}{\rm \cite{LJB01}}\label{Graph_Algorithm}
A SCT problem $L=\zug{H, P, C, \G}$ is \emph{not} size-change terminating iff
$S$, the closure of $\G$ under composition, contains an idempotent SCG graph $G
: f \rightarrow f$ that does \emph{not} contain an arc of the form $x \descarrow
x$. 
\end{theorem}

\section{Size-Change Termination and Ramsey-Based Containment}\label{Sect:SCT_vs._Ramsey}
The Ramsey-based test of Section \ref{Sect:Ramsey-Based_Universality} and the
\LJB algorithm of Section \ref{SCT} bear a remarkable similarity. In this
section we bridge the gap between the Ramsey-based universality test and the
\LJB algorithm, by demonstrating that the \LJB algorithm is a specialized
realization of the Ramsey-based containment test. This first requires developing
a Ramsey-based framework for \buchi containment testing.

\subsection{Ramsey-Based Containment with Supergraphs}\label{Sect:Supergraphs}

To test the containment of a \buchi automaton $\A$ in a \buchi automaton $\B$,
we could construct the complement of $\B$ using either the Ramsey-based or
rank-based construction, compute the intersection automaton of $\A$ and
$\overline{\B}$, and search this intersection automaton for a lasso. With
universality, however, we avoided directly constructing $\overline{\B}$ by
exploiting the structure of states in the Ramsey-based construction (see Lemma
\ref{SVW_Checks}). We demonstrate a similar test for containment.

Consider two automata, $\A=\zug{\Sigma, Q_\A, Q^{in}_\A, \rho_\A, F_\A}$ and
$\B=\zug{\Sigma, Q_\B, Q^{in}_\B, \rho_\B, F_\B}$.  When testing the
universality of $\B$, any word not in $L(\B)$ is a sufficient counterexample.
To test $L(\A) \subseteq L(\B)$ we must restrict our search to the subset of
$\Sigma^\omega$ accepted by $\A$.  In Section \ref{Sect:Ramsey-Based_Universality},
we defined a set $\graph{Q}_\B$ of 0-1 arc-labeled graphs, whose elements provide a family of
$\omega$-languages that covers $\Sigma^\omega$ (see Lemma \ref{Proper_Covers}).  We now define a
set, $\superg{Q}_{\A,\B}$, which provides a family of $\omega$-languages covering $L(\A)$.

We first define $\arc{Q}_\A=Q_\A \times Q_\A$ to capture the connectivity in
$Q_\A$. An element $\arc{g}=\rzug{q,r} \in \arc{Q}_\A$ is a single arc asserting
the existence of a path in $\A$ from $q$ to $r$. With each arc we associate a
language, $L(\arc{g})$. 

\begin{defi}
Given $w \in \Sigma^+$, say that $w \in L(\rzug{q,r})$ iff there is a
path in $\A$ from $q$ to $r$ over $w$.
\end{defi}

Define $\superg{Q}_{\A,\B}$ as $\arc{Q}_\A \times \graph{Q}_\B$. The elements of
$\superg{Q}_{\A,\B}$, called {\em supergraphs}, are pairs consisting of an arc
from $\arc{Q}_\A$ and a graph from $\graph{Q}_\B$.  Each element simultaneously
captures all paths in $\B$ and a single path in $\A$.  The language
$L(\zug{\arc{g},\graph{g}})$ is then $L(\arc{g}) \cap L(\graph{g})$.  For
convenience, we implicitly take $\superg{g}=\zug{\arc{g}, \graph{g}}$, and say
$\zug{q,a,r} \in \superg{g}$ when $\zug{q,a,r} \in \graph{g}$. Since the
language of each graph consists of finite words, we employ the concatenation of
languages to characterize infinite runs. To do so, we first prove Lemma
\ref{C:X_Partitions}, which simplifies the concatenation of entire languages by
demonstrating an equivalence to the concatenation of arbitrary words from these
languages.

\begin{lem}\label{C:X_Partitions}
If $u \in L(\superg{g}),~v\in L(\superg{h}),~uv\in L(\superg{k})$, and
$L(\arc{g})\cdot L(\arc{h}) \subseteq L(\arc{k})$, then $L(\superg{g})\cdot
L(\superg{h}) \subseteq L(\superg{k})$ 
\end{lem} 
\begin{proof}
Assume we have such an $u$ and $v$. We demonstrate every word $w \in L(\superg{g})
\cdot L(\superg{h})$  must be in $L(\superg{k})$. 
If we expand the premise, we obtain $w \in (L(\arc{g}) \cap L(\graph{g})) \cdot
(L(\arc{h}) \cap L(\graph{h}))$. This implies $w$ must be in $L(\arc{g})\cdot
L(\arc{h})$ and in $L(\graph{g}) \cdot L(\graph{h})$.  Next, we know that $u
\in L(\graph{g}),~ v\in L(\graph{h}),$ and $uv\in L(\graph{k})$. Thus by Lemma
\ref{X_Partitions}, $L(\graph{g})\cdot L(\graph{h}) \subseteq L(\graph{k})$, and
$w \in L(\graph{k})$. Along with the premise $L(\arc{g})\cdot L(\arc{h})
\subseteq L(\arc{k})$, we can now conclude $w \in L(\arc{k}) \cap L(\graph{k})$,
which is $L(\superg{k})$. 
\end{proof}

The languages $L(\superg{g}),~ \superg{g} \in \superg{Q}_{\A,\B}$, cover all
finite subwords of $L(\A)$. A subword of $L(\A)$ has at least one
path between two states in $Q_A$, and thus is in the language of an arc in
$\arc{Q}_\A$.  Furthermore, by Lemma \ref{X_Partitions} this word is described by some graph, and
the pair of the arc and the graph makes a supergraph. Unlike the case of graphs and
$\Sigma^+$, the languages of supergraphs do not form a partition of $L(\A)$: a
word might have multiple paths between states in $\A$, and so be described by
more than one arc in $\arc{Q}_\A$. With them we construct the finite family of
$\omega$-languages that cover $L(\A)$. Given $\superg{g},~ \superg{h} \in
\superg{Q}_{\A,\B}$, let $\Z{g}{h}$ be the $\omega$-language
$L(\superg{g})\cdot L(\superg{h})^\omega$. In analogy to Section
\ref{Sect:Ramsey-Based_Universality}, call $\Z{g}{h}$ \emph{proper}
if: (1) $\Z{g}{h}$ is non-empty; (2) $\arc{g}=\rzug{q,r}$ and
$\arc{h}=\rzug{r,r}$ where $q \in Q^{in}_\A$ and $r \in F_\A$; (3)
$L(\superg{g})\cdot L(\superg{h}) \subseteq L(\superg{g})$ and
$L(\superg{h})\cdot L(\superg{h}) \subseteq L(\superg{h})$.  Call a pair of
supergraphs $\zug{\superg{g}, \superg{h}}$ proper if $\Z{g}{h}$ is proper.  We
note that $\Z{g}{h}$ is non-empty if $L(\superg{g})$ and $L(\superg{h})$ are
non-empty, and that, by the second condition, every proper $\Z{g}{h}$ is
contained in $L(\A)$.

\begin{lem}\label{Supergraphs_Cover}
Let $\A$ and $\B$ be two \buchi automata, and $\superg{Q}_{\A,\B}$ the
corresponding set of supergraphs. $L(\A) = \bigcup\{\Z{g}{h}~|~,\superg{g},\superg{h} \in
\superg{Q}_{\A,\B},~\Z{g}{h}\text{ is proper}\}$.
\end{lem}
\begin{proof}
We extend the Ramsey argument of Lemma \ref{Proper_Covers} to supergraphs.

Consider an infinite word $w=\sigma_0\sigma_1...$ with an accepting run
$p=p_0p_1...$ in $\A$.  As $p$ is accepting, we know that $p_0 \in Q^{in}_\A$
and $p_i \in F_\A$ for infinitely many $i$.  Since $F_\A$ is finite, at least
one accepting state $q$ must appear infinitely often. Let $D \subseteq \natnum$
be the set of indexes $i$ such that $p_i=q$.

We pause to observe that, by the definition of the languages of arcs, for every
$i \in D$ the word $\sigma_0...\sigma_{i-1}$ is in $L(\rzug{p_0,q})$, and for every $i,j
\in D$, $i<j$, the word $\sigma_i...\sigma_{j-1} \in L(\rzug{q,q})$. Every language
$\Z{g}{h}$ where $\arc{g}=\rzug{p_0,q}$ and $\arc{g}=\rzug{q,q}$ thus satisfies
the second requirement of properness.

In addition to restricting our attention the subset of nodes where $p_i=q$, 
we further partition $D$ into $k=3^{n^2}$ sets $D_1,...,D_k$ based on the prefix
of $w$ until that point, where there is a $D_l$ associated with each possible
graph $\graph{g}_l$. By Lemma \ref{X_Partitions}, every finite word is in the
language of some graph $\graph{g}$.  Say that $i \in D_l$ iff $\sigma_0...\sigma_{i-1} \in
L(\graph{g}_l)$.  As $k$ is finite, for some $m$ $D_m$ must be infinite.  Let
$\graph{g}=\graph{g}_m$.

Similarly, by Lemma \ref{X_Partitions} we can use the word $w$ to define a
partition of all unordered \emph{pairs} of elements from $D_m$. This partition consists
of $k$ sets $C_1,...C_k$, such that $(i,j) \in C_l$ iff $\sigma_i...\sigma_{j-1} \in
L(\graph{g}_l)$. Without loss of generality, for $(i,j) \in C_l$, assume $i < j$.  Ramsey's Theorem
tells us that, given such a partition, there exists an infinite subset $\{i_1,i_2,...\}$ of $D_m$
and a $C_n$ such that $(i_j,i_k) \in C_n$ for all pairs of distinct elements $i_j,i_k$. 

This is precisely to say there is a graph $\graph{h}$ so that, for every $(i_j,
i_k) \in C_n$, it holds that  $\sigma_{i_j}...\sigma_{i_k-1} \in L(\graph{h})$.
$C_n$ thus partitions the word $w$ into $$w_1=\sigma_0...\sigma_{i_1-1},~ ~ ~ ~ 
w_2=\sigma_{i_1}...\sigma_{i_2-1},~ ~ ~ ~ 
w_3=\sigma_{i_2}...\sigma_{i_3-1},~ ~ ~ ~ ...,$$ 
such that $w_1 \in L(\graph{g})$ and $w_i \in L(\graph{h})$ for $i>1$.
Let $\superg{g}=\zug{\rzug{p_0,q},\graph{g}}$ and let
$\superg{h}=\zug{\rzug{q,q},\graph{h}}$. By the above partition of $w$, we know
that $w \in \Z{g}{h}$.

We now show that $\Z{g}{h}$ is proper.  First, as $w \in \Z{g}{h}$, we know $\Z{g}{h}$ is non-empty.
Second, as noted above, the second requirement is satisfied by the arcs $\rzug{p_0,q}$ and
$\rzug{q,q}$. Finally, we demonstrate the third condition holds. As $\sigma_0...\sigma_{i-1}
\in L(\graph{g})$ for every $i \in C_n$, we have that $w_1w_2 \in L(\graph{g})$.  Both $w_1$ and
$w_1w_2$ are in $L(\rzug{p_0,q})$ and so $w_1,\, w_1w_2 \in L(\superg{g})$.  By the definition of
the language of arcs, $L(\rzug{p_0,q})\cdot L(\rzug{q,q})\subseteq L(\rzug{p_0,q})$. Thus by Lemma
$\ref{C:X_Partitions}$, we can conclude that $L(\superg{g})\cdot L(\superg{h})\subseteq
L(\superg{g})$. Next observe that as $\sigma_{i}...\sigma_{j-1} \in L(\graph{h})$ for every pair $i, j \in C_n$,
we have that $w_2w_3 \in L(\graph{h})$. As $w_2,\,w_2w_3$ are both in $\rzug{q,q}$, it holds that
$w_2,\,w_2w_3 \in L(\superg{h})$.  By the definition of the language of arcs, $L(\rzug{q,q})\cdot
L(\rzug{q,q})\subseteq L(\rzug{q,q})$.  By Lemma \ref{C:X_Partitions} we can now conclude
$L(\superg{h})\cdot L(\superg{h})\subseteq L(\superg{h})$. Therefore $\Z{g}{h}$ is a proper language
containing $w$.
\end{proof}

\begin{lem}\label{SVW_Checks_Containment} 
Let $\A$ and $\B$ be two \buchi automata, and $\superg{Q}_{\A,\B}$ 
the corresponding set of supergraphs.  
\begin{enumerate}[\em(1)]
\item\label{Ccov:b} For all proper $\Z{g}{h}$, either $\Z{g}{h} \cap L(\B)=\emptyset$ or $\Z{g}{h} \subseteq L(\B)$.
\item\label{Ccov:c} $L(\A) \subseteq L(\B)$ iff every proper language $\Z{g}{h} \subseteq L(\B)$.
\item\label{Ccov:d} Let $\superg{g}, \superg{h}$ be two supergraphs such that
$\Z{g}{h}$ is proper. $\Z{g}{h} \subseteq L(\B)$ iff there exists $q \in
Q^{in}_\B,~ r \in Q_\B,~ a \in \{0,1\}$ such that $\zug{q,a,r} \in \superg{g}$
and $\zug{r,1,r} \in \superg{h}$.
\end{enumerate}
\end{lem} 
\begin{proof} Given two supergraphs $\superg{g}=\zug{\arc{g},\graph{g}}$ and
$\superg{h}=\zug{\arc{h},\graph{h}}$, recall that $\Y{g}{h}$ is the
$\omega$-language $L(\graph{g}) \cdot L(\graph{h})^\omega$. Further note that
$L(\superg{g}) \subseteq L(\graph{g})$ and $L(\superg{h}) \subseteq
L(\graph{h})$, and therefore $\Z{g}{h} \subseteq \Y{g}{h}$. 

\standout{\ref{Ccov:b}:} Consider two supergraphs $\superg{g}, \superg{h}$.
By Lemma \ref{Proper_Disjoint} either $\Y{g}{h} \cap L(\B)=\emptyset$ or
$\Y{g}{h} \subseteq L(\B)$.  Since $\Z{g}{h} \subseteq \Y{g}{h}$, it holds that
$\Z{g}{h} \cap L(\B) =\emptyset$ or $\Z{g}{h} \subseteq L(\B)$.

\standout{\ref{Ccov:c}:} Immediate from Lemma \ref{Supergraphs_Cover} and clause \ref{Ccov:b}.

\standout{\ref{Ccov:d}:} 
By Lemma \ref{Proper_Disjoint} either $\Y{g}{h} \subseteq L(\B)$ or $\Y{g}{h} \cap
L(\B) = \emptyset$. By Lemma \ref{SVW_Checks} $\Y{g}{h} \subseteq L(\B)$ iff a
$q, r$ and $a$ exist such that $\zug{q,a,r} \in \graph{g}$ and $\zug{r,1,r} \in
\graph{h}$. Since $\Z{g}{h} \subseteq \Y{g}{h}$,  $\Z{g}{h} \subseteq L(\B)$ iff
such a $q, r$ and $a$ exist.
\end{proof} 

In an analogous fashion to Section \ref{Sect:Ramsey-Based_Universality}, we can
use supergraphs to test the containment of two automata, $\A$ and $\B$. Search
all pairs of supergraphs, $\superg{g}, \superg{h} \in \superg{Q}_{\A,\B}$ for a
pair that is both proper and for which there does not exist a $q \in Q^{in}_\B$,
$r \in Q_\B, a \in \{0,1\}$ such that $\zug{q,a,r} \in \superg{g}$ and
$\zug{r,1,r} \in \superg{h}$.  Such a pair is a counterexample to containment.
If no such pair exists, then $L(\A) \subseteq L(\B)$.

\subsection{Composition of Supergraphs}

Employing supergraphs to test containment faces difficulty on two fronts. First, the number of
supergraphs is very large. Second, verifying properness requires checking language nonemptiness and
containment: PSPACE-hard problems. To address these problems we construct only
supergraphs with non-empty languages.  Borrowing the notion of composition from Section \ref{SCT}
allows us to use exponential space to compute exactly the needed supergraphs. Along the way we
develop a polynomial-time test for the containment of supergraph languages.  Our plan is to start
with graphs corresponding to single letters and compose them until we reach closure. The resulting
subset of $\superg{Q}_{\A,\B}$, written $\superg{Q}_{\A,\B}^f$, contains exactly the supergraphs
with non-empty languages. In addition to removing the need to check for emptiness, composition
allows us to test the sole remaining aspect of properness, language containment, in time polynomial
in the size of the supergraphs.  We begin by defining the composition of simple graphs.

\begin{defi}
Given two graphs $\graph{g}$ and $\graph{h}$ define their
composition, written as $\graph{g};\graph{h}$, as the graph  
\begin{eqnarray*}
&&\{\zug{q,1,r}~|~q,r,s \in Q_\B,\,\zug{q,b,s} \in \graph{g},\,\zug{s,c,r} \in
\graph{h},\,b=1\text{ or }c=1\}\\
&\cup& \{\zug{q,0,r}~|~q,r,s \in Q_\B,\,\zug{q,0,s} \in \graph{g},\,\zug{s,0,r} \in
\graph{h},\text{ and}\\
&&\text{ for all } t \in Q_\B,\, b,c \in \{0,1\} \text{ if } \zug{q,a,t} \in
\graph{g} \text{ and }~ \zug{t,b,r} \in \graph{h}\text{ then }a=b=0\}
\end{eqnarray*}
\end{defi}

\begin{figure}[t]
\begin{center}
{\includegraphics[clip=true, trim = 1.1in 7.3in 1.3in 1.1in,
height=0.22\linewidth]{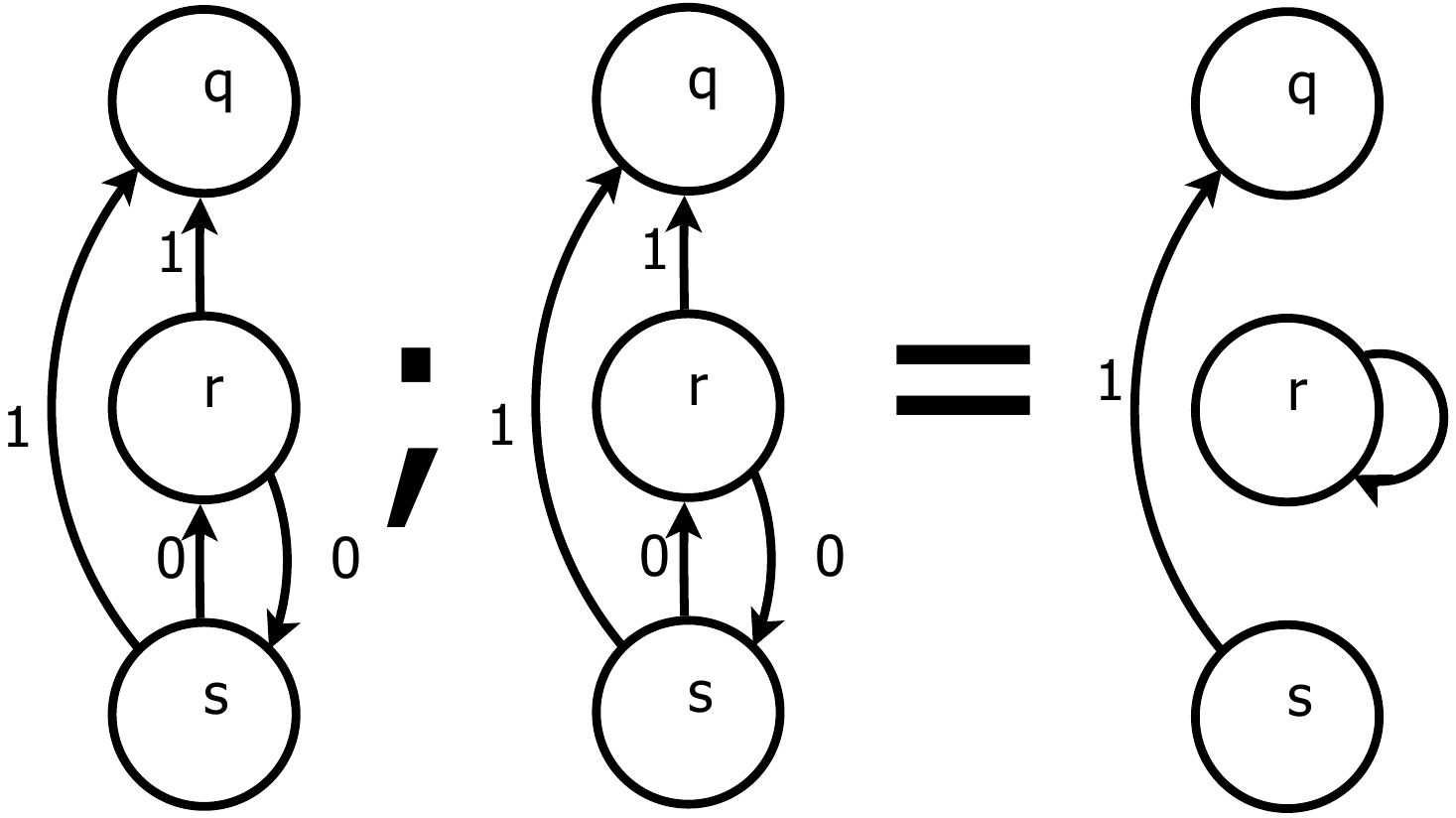}}
\end{center}
\caption{The composition of a graph with itself.}\label{fig:svw-comp}
\end{figure}

\begin{exa}
Figure \ref{fig:svw-comp} shows the composition of a simple graph with itself.
Figure \ref{fig:svw-graph} is also illustrative, as the third graph is
the composition of the first two.
\end{exa}

We can then define the composition of two supergraphs
$\superg{g}=\zug{\rzug{q,r},\graph{g}}$ and $\superg{h}=\zug{\rzug{r,s},\graph{h}}$, written
$\superg{g};\superg{h}$, as the supergraph $\zug{\rzug{q,s},\graph{g};\graph{h}}$.  To generate
exactly the set of supergraphs with non-empty languages, we start with supergraphs describing single
letters. For a containment problem $L(\A) \subseteq L(\B)$, define the subset of $\superg{Q}_{\A,\B}$
corresponding to single letters to be $\superg{Q}_{\A,\B}^{\init}=\{\superg{g}~|~\superg{g} \in
\superg{Q}_{\A,\B},~ a \in\Sigma,~ a \in L(\superg{g})\}$.  For completeness, we present a
constructive definition of $\superg{Q}_{\A,\B}^{\init}$.

\begin{defi}\label{Def:Buchi_to_Graphs}
\mbox{  }
\begin{tabbing}
\qquad$\superg{Q}_{\A,\B}^{\init} = 
{\big\{}\zug{\rzug{q,r}, \graph{g}}\mid~$\=$q\in Q_\A, r \in \rho_\A(q, a),~ a\in\Sigma,$\\
\>$\graph{g} =~$\=$\{\zug{q',0,r'} \mid q' \in Q_\B \setminus F_\B ,~ r' \in (\rho_\B(q', a) \setminus F_\B)\}~\cup$\\
\>\>$\{\zug{q',1,r'}\mid q' \in Q_\B,~ r' \in \rho_\B(q', a),~q' \text{ or } r' \in  F_\B)\}{\big\}}$
\end{tabbing}
\end{defi}

We then define $\superg{Q}_{\A,\B}^f$ to be the closure of $\superg{Q}_{\A,\B}^\init$ under
composition.   Algorithm \\\DGS, which we prove correct below, employs composition to check the
containment of two automata. It first generates the set of initial supergraphs, and then computes
the closure of this set under composition. Along the way it tests properness by using composition.
Every time it encounters a proper pair of supergraphs, it either verifies that a satisfying pair
of arcs exist, or halts with a counterexample to containment. We call this search the
\emph{double-graph search}. 

\begin{algorithm}[htbp]
\caption{\FuncSty{DoubleGraphSearch($\A$,$\B$)}}
\label{Alg:DoubleGraphSearch}
\DontPrintSemicolon
\KwData{Two \buchi automata, $\A$ and $\B$.}
\KwResult{Whether $L(\A)$ is contained in $L(\B)$.}
  Initialize $\superg{Q}^f \Leftarrow \superg{Q}_{\A,\B}^\init$\;
  \Repeat{$\superg{Q}^f$ reaches closure}
  {
  \For{\emph{\KwSty{all}} pairs $\superg{g},\superg{h} \in \superg{Q}^f$ where $\arc{g}=\zug{q,r}$ and $\arc{h}=\zug{r,s}$}
  {
      Add $\superg{g};\superg{h}$ to $\superg{Q}^f$\;
      \If{$q \in Q_\A^{in}$, ~ $r \in F_\A$, ~ $s=r$, ~ $\graph{g};\graph{h}=\graph{g}$ and $\graph{h};\graph{h}=\graph{h}$}
      {
          \If{there do not exist $\zug{q,a,r} \in \superg{g}$ and $\zug{r,1,r} \in \superg{h}$ where $q \in Q^{in}_\B$}
          {
			        \Return{Not Contained}
          }
      }
  }
  }
	\Return{Contained}
\end{algorithm}

To begin proving our algorithm correct, we link composition and the
concatenation of languages, first for simple graphs and then for
supergraphs. 

\begin{lem}\label{Univ_Concat_Valid}
For every two graphs $\graph{g}$ and $\graph{h}$, it holds that $L(\graph{g}) \cdot L(\graph{h})\subseteq L(\graph{g};\graph{h})$.
\end{lem} 
\begin{proof}
Consider two words $w_1 \in L(\graph{g}),~ w_2 \in L(\graph{h})$.  By
Definition \ref{X_Describes}, to prove $w_1w_2 \in L(\graph{g};\graph{h})$ we
must show that for every $q,r \in Q$: both (1) $\zug{q,a,r} \in \graph{g};\graph{h}$
iff there is a path from $q$ to $r$ over $w_1w_2$, and (2) that $a=1$ iff there
is an accepting path.

If an arc $\zug{q,a,r} \in \graph{g};\graph{h}$ exists, then there is an $s
\in Q$ such that $\zug{q,b,s} \in \graph{g}$ and $\zug{s,c,r} \in
\graph{h}$. By Definition \ref{X_Describes}, this implies the existence of a
path $x_1s$ from $q$ to $s$ over $w_1$, and a path $sx_2$ from $s$ to $r$ over
$w_2$. Thus $x_1sx_2$ is a path from $q$ to $r$ over $w_1w_2$. 

If $a$ is $1$, then either $b$ or $c$ must be 1. By Definition
\ref{X_Describes}, $b$ (resp., $c$) is 1 iff there is an accepting path $x'_1s$
(resp., $sx'_2$) over $w_1$ (resp.,$w_2$) from $q$ to $s$ (resp., $s$ to $r$).
In this case $x'_1sx_2$ (resp., $x_1sx'_2$) is an accepting path in $\B$ from
$q$ to $r$ over $w_1w_2$. 

Symmetrically, if there is a path $x$ from $q$ to $r$ over $w_1w_2$, then
after reading $w_1$ we are in some state $s$ and have split $x$ into
$x_1sx_2$, so that $x_1s$ is a path from $q$ to $s$ and $sx_2$ a path from
$s$ to $r$.  Thus by Definition \ref{X_Describes} $\zug{q,b,s} \in \graph{g}$,
$\zug{s,c,r} \in \graph{h}$, and $\zug{q,a,s} \in \graph{g};\graph{h}$.

Furthermore, if there is an accepting path from $q$ to $r$ over $w_1w_2$,
then after reading $w_1$ we are in some state $s$ and have split the path into
$x_1sx_2$, so that $x_1s$ is a path from $q$ to $s$, and $sx_2$ a path from
$s$ to $r$.  Either $x_1s$ or $sx_2$ must be accepting, and thus by Definition
\ref{X_Describes} $\zug{q,b,s} \in \graph{g}$, $\zug{s,c,r} \in \graph{h}$, and
either $b$ or $c$ must be $1$. Therefore $a$ must be~$1$.
\end{proof}

\begin{lem}\label{Cont_Concat_Equiv_Subset}
Let $\superg{g}, \superg{h}, \superg{k}$ be supergraphs in
$\superg{Q}_{\A,\B}^f$ such that $\arc{g}=\rzug{q,r}$,~
$\arc{h}=\rzug{r,s}$, and $\arc{k}=\rzug{q,s}$. Then $\superg{g};\superg{h}
= \superg{k}$ iff $L(\superg{g})\cdot L(\superg{h})\subseteq L(\superg{k})$.
\end{lem} 
\begin{proof}
Assume $\superg{g};\superg{h} = \superg{k}$ as a premise.  This implies $\superg{k}
= \zug{\rzug{q,s}, \graph{g};\graph{h}})$. If either $L(\superg{g})$ or $L(\superg{h})$ are empty,
then $L(\superg{g})\cdot L(\superg{h})$ is empty and this direction holds trivially.  Otherwise,
take two words $u \in L(\superg{g}),~ v \in L(\superg{h})$.  By construction,  $u \in L(\rzug{q,r})$
and $v \in L(\rzug{r,s})$.  The definition of the languages of arcs therefore implies the existence
of a path from $q$ to $r$ over $u$ and a path from $r$ to $s$ over $v$. Thus $uv \in L(\rzug{q,s})$.
Similarly, $u \in L(\graph{g}),~ v \in L(\graph{h})$, and Lemma \ref{Univ_Concat_Valid} implies that
$uv \in L(\graph{k})$.  Thus $uv$ is in $L(\zug{\rzug{q,r}, \graph{k}})$.  and by Lemma
\ref{C:X_Partitions} $L(\superg{g}) \cdot L(\superg{h})\subseteq L(\superg{k})$.

In the other direction, if $L(\superg{g})\cdot L(\superg{h})\subseteq L(\superg{k})$, we show that
$\superg{g};\superg{h} = \superg{k}$.  By definition, $\superg{g};\superg{h}$ is $\zug{\rzug{q,s},
\graph{g};\graph{h}}$. As $\superg{g}, \superg{h} \in \superg{Q}_{\A,\B}^f$, they are the
composition of a finite number of graphs from $\superg{Q}_{\A,\B}^\init$. The above direction then
demonstrates that they are non-empty, and there is a word $w \in L(\superg{g}) \cdot L(\superg{h})$.
This expands to $w \in (L(\arc{g}) \cap L(\graph{g})) \cdot (L(\arc{h}) \cap L(\graph{h}))$, which
implies $w \in L(\graph{g}) \cdot L(\graph{h})$. By Lemma \ref{Univ_Concat_Valid}, $w$ is then in
$L(\graph{g};\graph{h})$. Since, by Lemma \ref{X_Partitions}, $w$ is the language of exactly one
graph, we have that $\graph{g};\graph{h} =\graph{k}$, which proves  $\superg{g};\superg{h} =
\superg{k}$.
\end{proof}

Lemma \ref{Cont_Concat_Equiv_Subset} provides the polynomial time test for
properness employed in Algorithm \linebreak[4]\DGS.  Namely, given
two supergraphs $\superg{g}=\zug{\rzug{q,r},\graph{g}}$ and
$\superg{h}=\zug{\rzug{r,r},\graph{h}}$ from $\superg{Q}_{\A,\B}^f$,
the pair $\zug{\superg{g},\superg{h}}$ is proper exactly when $q \in Q_\A^{in}$,
~ $r \in F_\A$, ~ $\graph{g};\graph{h}=\graph{g}$ and
$\graph{h};\graph{h}=\graph{h}$.  We now provide the final piece of our
puzzle: proving that the closure of $\superg{Q}_{\A,\B}^{\init}$ under
composition contains every non-empty supergraph.

\begin{lem}\label{Cont_Concat_Complete}
For two \buchi automata $\A$ and $\B$, every $\superg{h} \in
\superg{Q}_{\A,\B}$, where $L(\superg{h})\neq\emptyset$, is in 
$\superg{Q}_{\A,\B}^f$.
 \end{lem} 
\begin{proof}
Let $\superg{h}=\zug{\rzug{q,r},\graph{h}}$ where $L(\superg{h}) \neq
\emptyset$. Then there is at least one word$w =\sigma_0...\sigma_{n-1} \in
L(\superg{h})$, which is to say $w \in L(\rzug{q,r}) \cap L(\graph{h})$.  By
the definition of the languages of arcs, there is a path $p=p_0...p_n$ in $\A$
over $w$ such that $p_0=q$ and $p_n=r$.

Define $\graph{g}_{\sigma_i}$ to be the graph in $\graph{Q}_\B^{\init}$ containing
$\sigma_i$. Let $\superg{g}_{\sigma_i}$ be $\zug{\rzug{p_i,p_{i+1}}, \graph{g}_{\sigma_i}}$,
and let $\superg{g}_w$ be $\superg{g}_{\sigma_0}; \superg{g}_{\sigma_1};
...;\superg{g}_{\sigma_{n-1}}$.  Note that each $\superg{g}_{\sigma_i} \in
\superg{Q}_{\A,\B}^\init$. By Lemma \ref{Cont_Concat_Equiv_Subset} $w \in
\graph{g}_w$. By Lemma \ref{X_Partitions}, $w$ is in only one graph and
$\graph{g}_w=\graph{h}$. By construction, $\arc{g}_w=\rzug{q,r}$.  Therefore
$\zug{\arc{g}_w,\graph{g}_w} =\zug{\rzug{q,r},\graph{h}} =\superg{h}$, and
$\superg{h}$ is in the closure of $\superg{Q}_{\A,\B}^{\init}$ under
composition.  
\end{proof}

We can now show the correctness of Algorithm \DGS, using Lemma \ref{Cont_Concat_Equiv_Subset} to
justify testing properness with composition, and Lemma \ref{Composition_Containment_Complete} below
to justify the correctness and completeness of our search for a counterexample.

 \begin{lem}\label{Composition_Containment_Complete}
Let $\A$ and $\B$ be two \buchi automata. $L(\A)$ is \emph{not} contained in
$L(\B)$ iff $\superg{Q}^f_{\A,\B}$ contains a pair of supergraphs $\superg{g}$,
$\superg{h}$ such that $\zug{\superg{g},\superg{h}}$ is proper and there do
\emph{not} exist arcs $\zug{q,a,r} \in \superg{g}$ and $\zug{r,1,r} \in
\superg{h}$, $q \in Q^{in}_\B$.
\end{lem} 
\begin{proof}
As all proper graphs are non-empty, this follows from parts (2) and (3) of
Lemma \ref{SVW_Checks_Containment} and Lemma \ref{Cont_Concat_Complete}.
\end{proof}

\begin{theorem}\label{Composition_Algorithm}
For every  two \buchi automata $\A$ and $\B$, it holds that $L(\A) \subseteq
L(\B)$ iff \linebreak[4]\DGS{$\A$,$\B$} returns {Contained}.
\end{theorem}
\begin{proof}
By Lemma \ref{Cont_Concat_Equiv_Subset}, testing for composition is equivalent
to testing for language containment, and the outer conditional in Algorithm \DGS
holds only for proper pairs of supergraphs. By Lemma
\ref{Composition_Containment_Complete}, the inner conditional checks if a proper
pair of supergraphs is a counterexample, and if no such proper pair in
$\superg{Q}^f_{\A,\B}$ is a a counterexample then containment must hold.
\end{proof}

\subsection{Strongly Suffix Closed Languages}\label{Sect:Suffix_Closed}

Algorithm \DGS has much the same structure as
Algorithm \LJB. The most noticeable difference is that Algorithm
\linebreak[4]\DGS checks pairs of supergraphs, where Algorithm
\LJB checks only single size-change graphs. Indeed, Theorem
\ref{Graph_Algorithm} suggests that, for some languages, a cycle implies the
existence of a lasso. When proving containment of \buchi automata with such
languages, it is sufficient to search for a graph $\superg{h} \in
\superg{Q}_\B$, where $\superg{h};\superg{h}=\superg{h}$, with no arc
$\zug{r,1,r}$. This \emph{single-graph search} reduces the complexity of our algorithm
significantly. What enables this in size-change termination is the late-start
property: threads can begin at arbitrary points. We here define the class of automata
amenable to this optimization, first presenting the case for universality testing,
without proof, for clarity.

In size-change termination, the late-start property asserts that an accepting
cycle can start at an arbitrary point. Intuitively, this suggests that an arc
$\zug{r,1,r} \in \graph{h}$ might not need a matching prefix $\zug{q,a,r}$ in
some $\graph{g}$: the cycle can just start at $r$. In the context of
universality, we can apply this method when it is safe to add or remove
arbitrary prefixes of a word. To describe these languages we extend the standard
notion of \emph{suffix closure}.  A language $L$ is suffix closed when, for
every $w \in L$, every suffix of $w$ is in $L$.

\begin{defi}
A language $L$ is \emph{strongly suffix closed} if it is suffix closed and for
every $w \in L,~ w_1 \in \Sigma^+$, we have that $w_1w \in L$.
\end{defi}

\begin{lem}\label{Suffix_Closed_Universality}
Let $\B$ be an \buchi automaton where every state in $Q$ is reachable and
$L(\B)$ is strongly suffix closed. $\B$ is \emph{not} universal iff the set of
supergraphs with non-empty languages, $\graph{Q}^f_\B$, contains a graph
$\graph{h}$ such that $\graph{h};\graph{h}$ and $\graph{h}$ does \emph{not}
contain an arc of the form $\zug{r,1,r}$.
\end{lem} 

As an intuition for the correctness of Lemma \ref{Suffix_Closed_Universality},
note that the existence of an 1-labeled cyclic arc in $\graph{h}$ implies a
loop, that $Q$ being reachable implies a prefix can be prepended to this loop to
make a lasso, and that strong suffix closure allows us to swap this prefix for
the prefix of every other word that share this cycle. 

To extend this notion to handle containment questions $L_1 \subseteq L_2$, we
restrict our focus to words in $L_1$. Instead of requiring $L_2$ to be closed
under arbitrary prefixes, $L_2$ need only be closed under prefixes that keep the
word in $L_1$.

\begin{defi}
A language $L_2$ is \emph{strongly suffix closed with respect to $L_1$} when
$L_2$ is suffix closed and, for every $w \in L_1 \cap L_2,~ w_1 \in \Sigma^+$, if
$w_1w \in L_1$ then $w_1w \in L_2$.
\end{defi}

When checking the containment of $\A$ in $\B$ for the case when $L(\B)$ is
strongly suffix closed with respect to $L(\A)$, we can employ a the simplified
algorithm below. As in Algorithm \linebreak[4]\DGS, we search all supergraphs in
$\superg{Q}^f_{\A,\B}$. Rather than searching for a proper pair of supergraphs,
however, Algorithm \SGS searches for a single supergraph $\superg{h}$ where
$\superg{h};\superg{h}=\superg{h}$ that does not contain an arc of the form
$\zug{r,1,r}$. We call this search the
\emph{single-graph search}.

\begin{algorithm}[htp]
\caption{\FuncSty{SingleGraphSearch($\A$,$\B$)}}
\label{Alg:SingleGraphSearch}
\DontPrintSemicolon
\KwData{Two \buchi automata, $\A$ and $\B$.}
\emph{\KwSty{Require}} $Q_\A^{in}=Q_\A$, $Q_\B$ is reachable, and $L(\B)$ is strongly suffix closed w.r.t. $L(\A)$\;
\KwResult{Whether $L(\A)$ is contained in $L(\B)$.}
  Initialize $\superg{Q}^f \Leftarrow \superg{Q}_{\A,\B}^\init$\;
  \Repeat{$\superg{Q}^f$ reaches closure}
  {
  \For{\emph{\KwSty{all}} pairs $\superg{g},\superg{h} \in \superg{Q}^f$ where $\arc{g}=\zug{q,r}$ and $\arc{h}=\zug{r,s}$}
  {
	  $\superg{k} \Leftarrow \superg{g};\superg{h}$\;
    Add $\superg{k}$ to $\superg{Q}^f$\;
    \If{$q \in F_\A$, $q=s$, and $\superg{k};\superg{k}=\superg{k}$}
    {
       \If{there does not exist an arc $\zug{r,1,r} \in \superg{k}$}
       {
			    \Return{Not Contained}
       }
    }
  }
  }
	\Return{Contained}
\end{algorithm}

We now prove Algorithm \SGS correct. Theorem \ref{Suffix_Closed_Containment}
demonstrates that, under the requirements specified, the presence of a
single-graph counterexample refutes containment, and the absence of such a
supergraph proves containment.  

\begin{theorem}\label{Suffix_Closed_Containment}
Let $\A$ and $\B$ be two \buchi automata where $Q^{in}_\A=Q_\A$, every state in
$Q_\B$ is reachable, and $L(\B)$ is strongly suffix closed with respect to
$L(\A)$. Then $L(\A)$ is \emph{not} contained in $L(\B)$ iff
$\superg{Q}_{\A,\B}^f$ contains a supergraph
$\superg{h}=\zug{\rzug{s,s},\graph{h}}$ such that $s \in F_\A$,
$\superg{h};\superg{h} = \superg{h}$, and $\superg{h}$ does \emph{not} contain an
arc $\zug{r,1,r}$.
\end{theorem} 
\begin{proof}
In one direction, assume $\superg{Q}_{\A,\B}^f$ contains a supergraph
$\superg{h}=\zug{\rzug{s,s},\graph{h}}$ where $s \in F_\A$,\linebreak[3]
$\superg{h};\superg{h} = \superg{h}$, and there is no arc $\zug{r,1,r} \in
\superg{h}$.  We show that $\Z{h}{h}$ is a proper language not contained in
$L(\B)$.  As $\superg{h} \in \superg{Q}_{\A,\B}^f$, we know $L(\superg{h})$ is
not empty, implying $\Z{h}{h}$ is non-empty. As $Q_\A=Q^{in}_\A$, it holds that $s \in
Q^{in}_\A$.  By Lemma \ref{Cont_Concat_Equiv_Subset}, the premise
$\superg{h};\superg{h} =\superg{h}$ implies $L(\superg{h})\cdot
L(\superg{h})\subseteq L(\superg{h})$, and $\Z{h}{h}$ is proper. Finally, as
there is no $\zug{r,1,r} \in \superg{h}$, by Theorem \ref{Composition_Algorithm},
$\Z{h}{h} \not \in L(B)$, and $\Z{h}{h}$ is a counterexample to $L(\A) \subseteq L(\B)$.

In the opposite direction, assume the premise that $\superg{Q}_{\A,\B}^f$ does
not contain a supergraph $\superg{h}=\zug{\rzug{s,s},\graph{h}}$ where $s \in
F_\A$, $\superg{h};\superg{h}=\superg{h}$, and there is no arc $\zug{r,1,r} \in
\superg{h}$.  We prove that every word $w \in L(\A)$ is also in $L(\B)$.  Take a
word $w \in L(\A)$.  By Lemma \ref{Supergraphs_Cover}, $w$ is in some proper
language $\Z{g}{h}$ and can be broken into $w_1w_2$ where $w_1 \in
L(\superg{g}),~ w_2 \in L(\superg{h})^\omega$. 

Because $\Z{g}{h}$ is proper, Lemma \ref{Cont_Concat_Equiv_Subset} implies
$\superg{h}=\zug{\rzug{s,s},\graph{h}}$ where $s \in F_\A$ and
$\superg{h};\superg{h}=\superg{h}$. This, along with our premise, implies
$\superg{h}$ contains an arc $\zug{r,1,r}$.  Since all states in $Q_\B$ are
reachable, there is $q \in Q^{in}_\B$ and $u \in \Sigma^+$ with a path in $\B$
from $q$ to $r$ over $u$. By Lemma \ref{SVW_Checks}, this implies $uw_2$ is
accepted by $\B$.  For $L(\B)$ to
strongly suffix closed with respect to $L(\A)$, it must be suffix closed.
Therefore $w_2 \in L(\B)$. Now we move to $L(\A)$, and note that the premise
$Q^{in}_\A=Q_\A$ implies $L(\A)$ is suffix closed. Thus the fact that $w_1w_2
\in L(\A)$ implies $w_2 \in L(\A)$.  Since $L(\B)$ is strongly suffix closed
with respect to $L(\A)$, and $w_2 \in L(\B)$, it must be that $w_1w_2 \in
L(\B)$. 
\end{proof}

\subsection{From Ramsey-Based Containment to Size-Change Termination}\label{Sect:SCT_Is_Ramsey}

We now delve into the connection between the \LJB algorithm for
size-change termination \linebreak[2]and the single-graph search algorithm for \buchi
containment. We will show that \linebreak[2]Algorithm \LJB is a specialized
realization of Algorithm \SGS. Given an\linebreak[2] SCT problem $L$,
size-change graphs in \LJB{$L$} are direct analogues of supergraphs in
\linebreak[2]\SGS{$\A_{Flow(L)}$, $\A_{Desc(L)}$}. For convenience, take $L=\zug{H, P, C,
\G}$, \linebreak[2]$\A_{Flow(L)}= \zug{C, Q_{Fl}, Q_{Fl}^{in}, \rho_{Fl}, F_{Fl}}$, and
$\A_{Desc(L)} = \zug{C, Q_{Ds}, Q_{Ds}^{in}, \rho_{Ds}, F_{Ds}}$.

We first show that $\A_{Flow(L)}$ and $\A_{Desc(L)}$ satisfy the preconditions 
of Algorithm \linebreak[4]\SGS: that $Q^{in}_{Fl}=Q_{Fl}$; that every state in
$Q_{Ds}$ is reachable; and that $Desc(L)$ is strongly suffix closed with respect
to $Flow(L)$.  For the first and second requirement, it suffices to observe that
every state in both $\A_{Flow(L)}$ and $\A_{Desc(L)}$ is initial.\footnote{In
the original reduction, 1-labeled parameters may not have been reachable.} 

For the third, strong suffix closure is a direct consequence of the definition
of a thread: since a thread can start at arbitrary points, it does not matter
what call path we use to reach that point. Adding a prefix to a call path cannot
cause that call path to become non-terminating. Thus the late-start property is
precisely $Desc(L)$ being strongly suffix closed with respect to $Flow(L)$, and
we can employ the single-graph search. 

Consider supergraphs in $\superg{Q}_{\A_{Flow(L)},\A_{Desc(L)}}$, from here
simply denoted by $\supergFD$.  The state space of $\A_{Flow(L)}$ is the set
of functions $H$, and the state space of $\A_{Desc(L)}$ is the union of $H$ and
$Q_p$, the set of all $\{0,1\}$-labeled parameters.  A supergraph in
$\supergFD$ thus comprises an arc $\rzug{q,r}$ in
$H$ and a $\{0,1\}$-labeled graph $\graph{g}$ over $H \cup Q_p$. The arc asserts
the existence of a call path from $q$ to $r$, and the graph $\graph{g}$ captures
the relevant information about corresponding paths in $\A_{Desc(L)}$. 

\begin{figure}[tb]
\begin{center}
{\includegraphics[clip=true, trim = 1.1in 8.6in 1.1in 1.1in, height=0.25\linewidth]{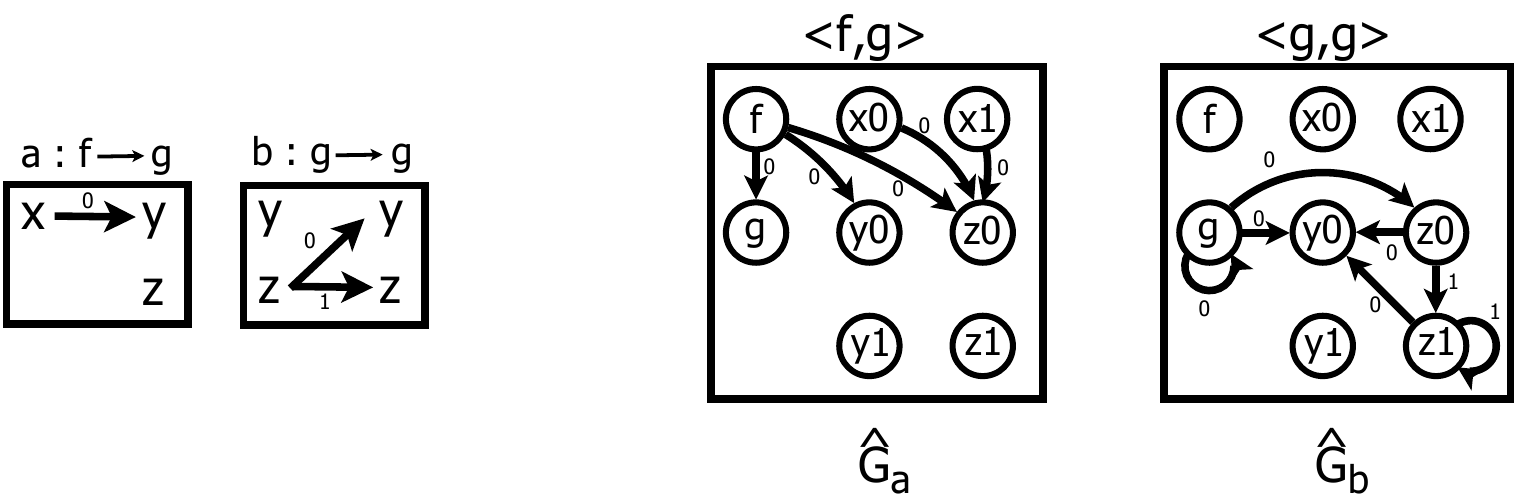}}
\end{center}
\caption{Size-Change Graphs vs. Supergraphs: The SCGs for call sites $a$ and $b$, from Figure \ref{fig:scg}, and corresponding
supergraphs for the characters $a$ and $b$, from $\superg{Q}^\init_{\A_{Flow(L)},\A_{Desc(L)}}$.}\label{fig:scg_vs_supergraphs}
\end{figure}

These supergraphs are almost the same as SCGs, $G : q \rightarrow r$ (See Figure
\ref{fig:scg_vs_supergraphs}).  Aside from notational differences, both contain an arc asserting the
existence of a call path between two functions, and a $\{0,1\}$-labeled graph.  There are nodes in
both graphs that correspond to parameters of functions,
and arcs between two such nodes describe a thread between the corresponding
parameters. The analogy falls short, however, on three points:
%\begin{asparaenum}
\begin{enumerate}[(1)]
\item\label{Dif:FNames} In SCGs, nodes are always parameters of
functions.  In supergraphs, nodes can be either parameters of functions or
function names. 
\item\label{Dif:Labels} In SCGs, nodes are unlabeled. In
supergraphs, nodes are labeled either~$0$~or~$1$. 
\item\label{Dif:Width} In an SCG, only the nodes corresponding to
the parameters of two specific functions are present. In a supergraph, nodes 
corresponding to every parameter of every function exist. 
\end{enumerate}
%\end{asparaenum}

Each difference is an opportunity to specialize the Ramsey-based containment
algorithm, Algorithm \SGS, by simplifying supergraphs.
When these specializations are taken together, we have Algorithm \LJB.

\begin{enumerate}[(1)]
\item[{\ref{Dif:FNames}}] No functions in $H$ are accepting for $\A_{Desc(L)}$, and
once we transition out of $H$ into $Q_p$ we can never return to $H$. Therefore
nodes $r$ corresponding to function names can never be part of a descending arc
$\zug{r,1,r}$. Since we only search for arcs of the form $\zug{r,1,r}$, we can
simplify supergraphs in $\supergFD$ by removing all nodes corresponding to
functions. 

\item[{\ref{Dif:Labels}}] The labels on parameters are the result of encoding a \buchi
edge acceptance condition in a \buchi state acceptance condition automaton, and
can be dropped from supergraphs with no loss of information.  Consider an arc
$\zug{\zug{f,a}, b, \zug{g,c}}$. If $b$ is 1, we know the corresponding thread
contains a descending arc. The value of $c$ tells us if the final arc in the
thread is descending, but \emph{which arc} is descending is irrelevant.  Thus it is
safe to simplify supergraphs in $\supergFD$ by removing labels on parameters.

\item[{\ref{Dif:Width}}] While all parameters have corresponding states in
$\A_{Desc(L)}$, each supergraph describes threads in a call sequence between two
particular functions.  There are no threads in this call sequence between
parameters of other functions, and so no supergraph with a non-empty language
has arcs between the parameters of other functions. We can thus simplify
supergraphs in $\supergFD$ by removing all nodes corresponding to parameters of
other functions.
\end{enumerate}

To formalize this notion of simplification, we first define, $\superg{G}_L$, the set of
simplified supergraphs and show that $\superg{G}_L$ is in
one-to-one correspondence with $S$, the closure of $\G$ under composition.

\begin{defi}\label{ReducedSgraph}
$\superg{G}_L = \{\zug{\rzug{f_1,f_2}, \graph{k}}~|~f_1,f_2 \in H,~
\graph{k} \subseteq 2^{P(f_1) \times \{0,1\} \times P(f_2)}\}$ 
\end{defi}

Say that \emph{$\zug{r,\graph{g}} \in \supergFD$
simplifies to $\zug{r,\graph{k}} \in \superg{G}_L$} when $\zug{q,b,r} \in
\graph{k}$ iff there exists $a, c \in \{0,1\}$ such that
$\zug{\rzug{q,a},b,\rzug{r,c}} \in \graph{g}$.  Let $\superg{G}_L^{\init}$ be
$\{\superg{k}~|~\superg{g} \in \supergFD^{\init},~
\superg{g} \text{ simplifies to } \superg{k}\}$, and $\superg{G}_L^f$ be the
closure of $\superg{G}_L^{\init}$ under composition. 

We can map SCGs directly to elements of $\superg{G}_L$.  Say $G : f_1
\rightarrow f_2 \equiv \zug{\rzug{f_1,f_2}, \graph{g}}$ when $q \vararrow{a} r
\in G$ iff $\zug{q,a,r} \in \graph{g}$. Note that the composition operations for
supergraphs of this form is identical to the composition of SCGs: if $G_1 \equiv
\superg{g}$ and $G_2 \equiv \superg{h}$, then $G_1;G_2 \equiv
\superg{g};\superg{h}$.  Therefore every element of $\supergFD^f$ simplifies to
some element of $\superg{G}_L^f$.

We now show that supergraphs whose languages contain single characters are in
one-to-one correspondence with $\G$, and that every idempotent element of
$\superg{G}_L^f$ contains an arc of the form$\zug{r,1,r}$ exactly when the
closure of $\G$ under composition does not contain a counterexample graph. 

\begin{lem}
Let $L=\zug{H, P, C, \G}$ be an SCT problem.
\begin{enumerate}[\em(1)]
\item The $\equiv$ relation is a one-to-one correspondence between
$\superg{G}^\init_L$ and $\G$
\item $L$ is \emph{not} size-change terminating iff $\superg{G}^f_L$ contains a
supergraph $\superg{k}$ such that 
$\superg{k};\superg{k}=\superg{k}$ and there does \emph{not}
exist an arc of the form $\zug{r,1,r}$ in $\superg{k}$.
\end{enumerate}
\end{lem}
\begin{proof}\  

{(1):}
Given a size-change graph $G \in \G$, we construct a unique supergraph
$\superg{k} \in \superg{G}^\init$ such that $\superg{k} \equiv G$.  Every
size-change graph $G : f_1 \rightarrow f_2 \in \G$ is the SCG for a call site
$c$ from $f_1$ to $f_2$.  This is a call sequence of length one. Thus there is a
$\superg{g} \in \superg{Q}_L^\init$ so that $c \in L(\superg{g})$ and $\arc{g} =
\zug{f_1,f_2}$.  We show that the simplification of $\superg{g}$ is equivalent
to $G$.  By the reduction of Definition \ref{LJB_Reduction} and the definition
of graphs in Definition \ref{X_Describes}, the arc
$\zug{\rzug{q,b},a,\rzug{r,c}} \in \superg{g}$, for some $b,c \in \{0,1\}$,
exactly when $q \vararrow{a} r \in G$.  The supergraph $\superg{g}$ simplifies
to some $\superg{k} \in \superg{G}_L$. By the definition of simplification,
$\zug{\rzug{q,b},a,\rzug{r,c}} \in \superg{g}$ exactly when $\zug{q,a,r} \in
\superg{k}$.  Thus $\superg{k} \equiv G : f_1 \rightarrow f_2$. 

In the other direction, $\superg{k} \in \superg{G}^\init$ iff there exists a
$\superg{g}=\zug{\rzug{f_1,f_2},\graph{g}} \in \supergFD^\init$ that simplifies
to $\superg{k}$. By Definition \ref{Def:Buchi_to_Graphs}, which defines
$\supergFD^\init$ and Definition \ref{LJB_Reduction}, $\superg{g}$ exists
because there is a call site $c$. This call site corresponds to a SCG $G : f_1
\rightarrow f_2$.  Analogously to the above, by Definition \ref{LJB_Reduction}
the arcs in $\graph{g}$ between parameters correspond to arcs in $G$:
$\zug{\rzug{q,b},a,\rzug{r,c}} \in \graph{g}$, for some $b,c \in \{0,1\}$,
exactly when $q \vararrow{a} r \in G$.  These are the only arcs that remain
after simplification, during which the labels are removed. Thus $\superg{k}
\equiv G : f_1 \rightarrow f_2$. 

{(2):}
By (1), $\G$ is in one-to-one correspondence with $\superg{G}_L^{\init}$ under
the $\equiv$ relation. Since composition of supergraphs and SCGs is identical,
$S$, the closure of $\G$ under composition, is in one-to-one correspondence with
the $\superg{G}_L^f$.  Claim (2) then follows from Theorem \ref{Graph_Algorithm}
and claim (1). 
\end{proof}

In conclusion, we can specialize the Ramsey-based containment algorithm
for\linebreak[1]
$L(\A_{Flow(L)})$ $\subseteq L(\A_{Desc(L)})$ in two ways.  First, by Theorem
\ref{Suffix_Closed_Containment} we know that $Flow(L) \subseteq Desc(L)$ if and
only if $\supergFD$ contains an idempotent graph
$\superg{g}=\superg{g};\superg{g}$ with no arc of the form $\zug{r,1,r}$. Thus
we can employ the single-graph search instead of the double-graph search.
Secondly, we can simplify supergraphs in $\supergFD$ by removing the labels on
nodes and keeping only nodes associated with appropriate parameters for the
source and target function.  The simplifications of supergraphs whose languages
contain single characters are in one-to-one corresponding with $\G$, the initial
set of SCGs.  As every state in $Flow(L)$ is accepting, every idempotent
supergraph can serve as a counterexample. Therefore $Desc(L) \subseteq Flow(L)$
if and only if the closure of the set of simplified supergraphs, which is in
one-to-one correspondence with $\G$, under composition does not contain an
idempotent supergraph with no arc of the form $\zug{r,1,r}$. This is precisely
Algorithm \LJB.

\section{Empirical Analysis}\label{Sect:Exps}
All the Ramsey-based algorithms presented in Section \ref{SCT} have worst-case
running times that are exponentially larger than those of the rank-based
algorithms. We now compare existing, Ramsey-based, SCT tools tools to a
rank-based \buchi containment solver on the domain of SCT problems.  To
facilitate a fair comparison, we briefly describe two improvements to the
algorithms presented above.  

\subsection{Towards an Empirical Comparison}\label{Sect:Prep}

First, in constructing the analogy between SCGs in the \LJB algorithm and
supergraphs in the Ramsey-based containment algorithm, we noticed that
supergraphs contain nodes for every parameter, while SCGs contain only
nodes corresponding to parameters of relevant functions.  These nodes are
states in $\A_{Desc(L)}$. While we can specialize the Ramsey-based test to avoid
them, \buchi containment solvers might suffer. These states duplicate
information. As we already know which functions each supergraph corresponds to,
there is no need for each node to be unique to a specific parameter.

\begin{figure}[tb]
\begin{center}
{\includegraphics[clip=true, trim = 1.1in 7.2in 1.1in 1.0in, height=0.3\linewidth]{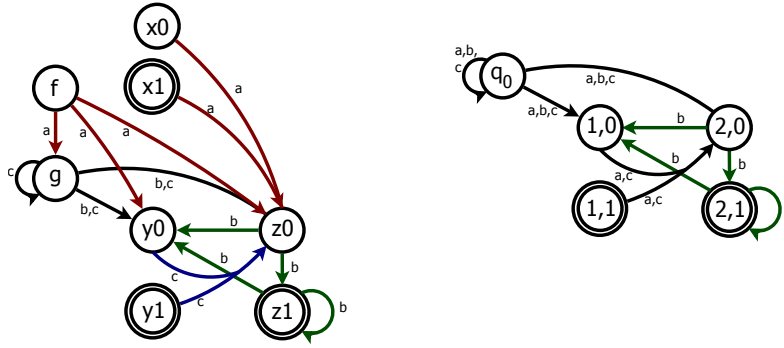}}
\end{center}
\caption{
$\A_{Desc(L)}$ (left), from the original reduction of Definition \ref{LJB_Reduction} , and 
$\A'_{Desc(L)}$ (right) from the reduction of Definition \ref{Desc_Reduction}. 
} \label{fig:desc_comp}
\end{figure}

These extra states emerge because $Desc(L)$ only accepts strings that are contained
in $Flow(L)$, and in doing so demands that parameters only be reached by
appropriate call paths.  But the behavior of $\A_{Desc(L)}$ on strings not in
$Flow(L)$ is irrelevant to the question of $Flow(L) \subseteq Desc(L)$, and we
can replace the names of parameters in $\A_{Desc(L)}$ with their location in the
parameter list.  Further, we can rely on $Flow(L)$ to verify the sequence of
function calls before our accepting thread and make do with a single waiting
state. As an example, see Figure \ref{fig:desc_comp}.

\begin{defi}\label{Desc_Reduction}
Given an SCT problem $L=\zug{H, P, C, \G}$ and a projection $Ar$ of all
parameters onto their positions $1..n$ in the parameter list, define:
\begin{eqnarray*}
\A'_{Desc(L)} &=& \zug{C, S\cup \{q_0\},S \cup \{q_0\},\rho_D,F}\\
&\textnormal{where }& S= \{1..n\} \times \{1, 0\}\\
&&    \rho_D(q_0, c)=\{q_0\} \cup \{\zug{Ar(x),0}~|~c:f_1 \rightarrow f_2,~ x \in P(f_2)\}\\
&&    \rho_D(\zug{h,a},c)=\{\zug{Ar(y),a'}~|~x\vararrow{a'}y\in\G_c,~ h=Ar(x)\}\\
&&    F = \{1..n\} \times \{1\}\\
\end{eqnarray*}
\end{defi}

\begin{lem}
$L(\A_{Flow(L)}) \subseteq L(\A_{Desc(L)})$ iff $L(\A_{Flow(L)}) \subseteq
L(\A'_{Desc(L)})$
\end{lem} 
\begin{proof}
The languages of $\A_{Desc(L)}$ and $\A'_{Desc(L)}$ are not the same. 
What we demonstrate is that for every word in $Flow(L)$, we can convert an
accepting run in one of $\A_{Desc(L)}$ or $\A'_{Desc(L)}$ into an accepting run
in the other.  Recall that the states of $\A_{Flow(L)}$ are functions $f \in H$.
States of $\A_{Desc(L)}$ are either elements of $H$ or elements of
$Q_p = \bigcup\nolimits_{f\in H} P(f) \times \{1, 0\}$, the set of labeled parameters.  
For convenience, given a pair $\zug{x,a} \in Q_p$, define $Ar(\zug{x,a})$ to be 
$\zug{Ar(x),a}$.

Consider an accepting run $r=r_0r_1...$ of $\A_{Desc(L)}$ over a
word $w$. Let $s=s_0s_1...$ be the sequence of states in $\A'_{Desc(L)}$ such
that when $r_i \in H$, $s_i=q_0$, and when $r_i \in Q_p$, $s_i=Ar(r_i)$. By
the definition of $\A'_{Desc(L)}$, $q_0$ always transitions to
$q_0$ and a transition between $r_i$ and $r_{i+1}$ implies a transition
between $Ar(r_i)$ and $Ar(r_{i+1})$. Therefore $s$ is a run of $\A'_{Desc(L)}$
over $w$. Furthermore, if $\zug{x,a}$ is an accepting state in $\A_{Desc(L)}$, $a=1$
and $\zug{Ar(x),a}$ is an accepting state $\A'_{Desc(L)}$. Thus, $s$ is an
accepting run of $\A'_{Desc(L)}$ over $w$.

Conversely, consider a word $w$ with an accepting run $r=r_0r_1...$ of $\A'_{Desc(L)}$ and
an accepting run $s=s_0s_1...$ of $\A_{Flow(L)}$.  We define an accepting
run $t=t_0t_1..$ of $\A_{Desc(L)}$ on $w$. Each $t_i$ depends on the
corresponding $r_i$ and $s_i$. If $r_i=q_0$ and $s_i=f$, then $t_i=f$.  If
$r_i=\zug{k,a}$  and $s_i=f$, then $t_i=\zug{x,a}$ where $x$ is the $k$th
parameter in $f$'s parameter list. 

For a call $c : f_1 \rightarrow f_2$, take two labeled parameters, $q$ a labeled
parameter of $f_1$ and $r$ a labeled parameter of $f_2$.  If
$\zug{Ar(q),c,Ar(r)}$ is a transition in $\A'_{Desc(L)}$, then $\zug{q, c, r}$
is a transition in $\A_{Desc(L)}$.  Therefore $t$ is a run of $\A_{Desc(L)}$ on
$w$.  Furthermore, note that $\zug{x,a} \in F_{\A_{Desc(L)}}$ and $\zug{Ar(x),a} \in
F_{\A_{Desc(L)}}$ iff $a=1$.  Therefore $t$ is an accepting run.
\end{proof}

Second, in \cite{BL07}, Ben-Amram and Lee present a polynomial approximation
of the \LJB algorithm for SCT. To facilitate a fair comparison, they optimize the
\LJB algorithm for SCT by using subsumption to remove certain SCGs when computing
the closure under composition. This suggests that the single-graph search of
Algorithm \SGS can also employ subsumption.  When
computing the closure of a set of graphs under compositions, we can ignore
elements when they are approximated by other elements.  Intuitively, a graph
$\graph{g}$ approximates another graph $\graph{h}$ when it is strictly harder to
find a 1-labeled sequence of arcs through $\graph{g}$ than through $\graph{h}$.
If we can replace $\graph{h}$ with $\graph{g}$ without losing 
arcs, we do not have to consider $\graph{h}$.  When the right arc can be found
in $\graph{g}$, then it also occurs in $\graph{h}$. On the other hand, when
$\graph{g}$ does not have a satisfying arc, then we already have a
counterexample.

\begin{figure}[tb]
\begin{center}
\includegraphics[width=0.3\linewidth]{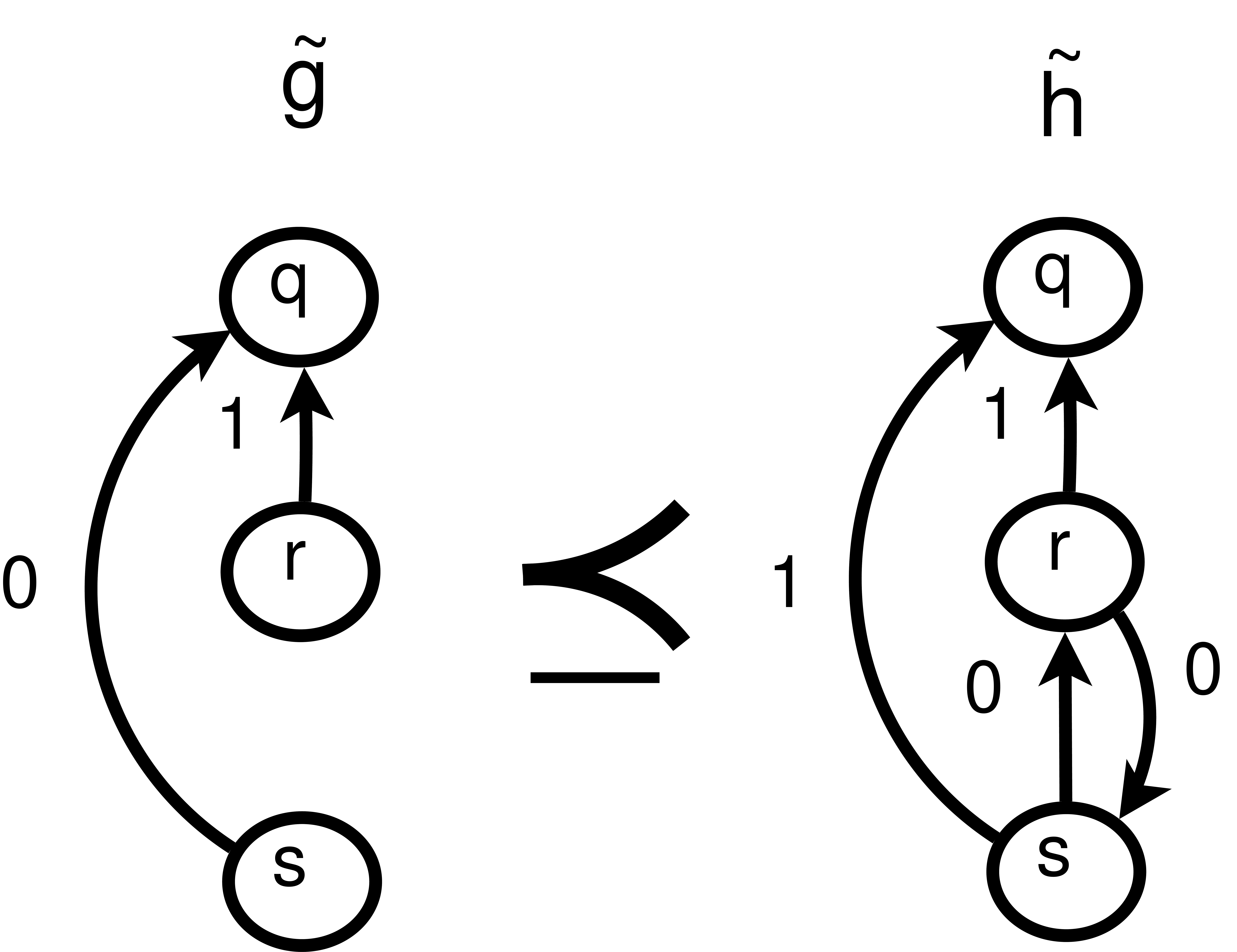}
\end{center}
\caption{Subsumption: two graphs $\graph{g}$ and $\graph{h}$, where
$\graph{g} \preceq \graph{h}$.}\label{fig:subsumption}
\label{fig:preceq}
\end{figure}

Formally, given two graphs $\graph{g}, \graph{h} \in \graph{Q}_\B$ say that $\graph{g}$
\emph{approximates} $\graph{h}$, written $\graph{g} \preceq \graph{h}$, when for every arc
$\zug{q,a,r} \in \graph{g}$ there is an arc $\zug{q,a',r} \in \graph{h}$, $a \leq a'$. An example is
provided in Figure \ref{fig:subsumption}. Note that approximation is a transitive relation.  In
order to safely employ approximation as a subsumption relation, Ben-Amram and Lee replace the search
for a single arc in idempotent graphs with a search for a strongly connected component in all
graphs. This was proven to be safe in \cite{Fog08}: when computing the closure of
$\graph{Q}_B^\init$ under composition, it is sufficient to store only maximal elements under this
relation. 

\subsection{Experimental Results}

All experiments were performed on a Dell Optiplex GX620 with a
single 1.7Ghz Intel Pentium 4 CPU and 512 MB. Each tool was given 3500
seconds, a little under one hour, to complete each task.

\standout{Tools:} The formal-verification community has implemented rank-based
tools in order to measure the scalability of various approaches.  The
programming-languages community has implemented several Ramsey-based SCT 
tools. We use the best-of-breed rank-based tool, \textbf{Mh},
developed by Doyen and Raskin \cite{DR07}, that leverages a subsumption relation
on ranks.  We expanded the Mh tool to handle \buchi containment problems with
arbitrary languages, thus implementing the full containment-checking algorithm
presented in their paper. 

We use two Ramsey-based tools. \textbf{SCTP} is a direct implementation of the
\LJB algorithm of Theorem \ref{Graph_Algorithm}, written in Haskell
\cite{Fred01}.  We have extended SCTP to reduce SCT problems to \buchi
containment problems, using either Definition \ref{LJB_Reduction} or
\ref{Desc_Reduction}. \textbf{sct/scp} is an optimized C implementation of the
SCT algorithm, which uses the subsumption relation of Section \ref{Sect:Prep}
\cite{BL07}. 

\standout{Problem Space:} Existing experiments on the practicality of SCT
solvers focus on examples extracted from the literature \cite{BL07}.  We combine
examples from a variety of sources
\cite{Daedalus,BL07,Fred01,Glen99,LJB01,SJ05,Wahl00}. The time spent reducing
SCT problems to \buchi automata never took longer than 0.1 seconds and was
dominated by I/O.  Thus this time was not counted\hide{Experimenting reveals
a statistically insignificant \emph{slowdown} when reading the SCT problem and
performing the reduction.}. We compared the performance of the rank-based Mh
solver on the derived \buchi containment problems to the performance of the
existing SCT tools on the original SCT problems. If an SCT problem was solved in
all incarnations and by all tools in less than 1 second, the problem was
discarded as uninteresting.  Unfortunately, of the 242 SCT problems derived from
the literature, only 5 prove to be interesting.

\standout{Experiment Results:} Table \ref{Table:SCT_Completion} compares the performance
of the rank-based Mh solver against the performance of the existing SCT tools,
displaying which problems each tool could solve, and the time taken to solve
them. Of the interesting problems, both SCTP and Mh could only complete 3.
On the other hand, sct/scp completed all of them, and had difficulty with only
one problem.

\begin{table}[!htp]
\vspace{0.17in}
\centering
\begin{tabular}{|l|c|c|c|}
\hline Problem & SCTP (s) & Mh (s) & sct/scp (s) \\
\hline ex04 \cite{BL07} & 1.58& Time Out & 1.39 \\
\hline ex05 \cite{BL07} & Time Out & Time Out & 227.7 \\
\hline ms \cite{Fred01} &  Time Out & 0.1 & 0.02 \\
\hline gexgcd \cite{Fred01} & 0.55 & 14.98 &  0.023 \\
\hline graphcolour2 \cite{Glen99} &  0.017 & 3.18 & 0.014 \\
\hline
\end{tabular}
\vspace{0.05in}
\caption{SCT problem completion time by tool.}\label{Table:SCT_Completion}
%\vspace{-0.3in}
\end{table}

The small problem space makes it difficult to draw firm conclusions, but it is
clear that Ramsey-based tools are comparable to rank-based tools on SCT
problems: the only tool able to solve all problems was Ramsey based. This is
surprising given the significant difference in worst-case complexity, and
motivates further exploration.

\section{Reverse-Determinism}\label{Sect:Rev-Det}
In the previous section, the theoretical gap in performance between Ramsey and
rank-based solutions was not reflected in empirical analysis. Upon further
investigation, it is revealed that a property of the domain of SCT problems is
responsible. Almost all problems, and every difficult problem, in this
experiment have SCGs whose nodes have an in-degree of at most 1. This
property was first observed by Ben-Amram and Lee in their analysis of SCT
complexity \cite{BL07}.  After showing how this property explains the
performance of Ramsey-based algorithms, we explore why this property emerges and
argue that it is a reasonable property for SCT problems to possess. Finally, we
improve the rank-based algorithm for problems with this property.

As stated above, all interesting SCGs in this experiment have nodes with at
most one incoming edge. In analogy to the corresponding property for automata,
we call this property of SCGs \emph{reverse-determinism}.  Say that a SCG is
reverse-deterministic if every parameter of $f_2$ has at most one incoming
edge.  Given a set of reverse-deterministic SCGs $\G$, we observe three
consequences.  First, a reverse-deterministic SCG can have no more than $n$
arcs: one entering each node. Second, there are only $2^{O(n \log n)}$
possible such combinations of $n$ arcs.  Third, the composition of two
reverse-deterministic SCGs is also reverse-deterministic.  Therefore every
element in the closure of $\G$ under composition is also reverse-deterministic.
These observations imply that the closure of $\G$ under composition contains at
most $2^{O(n \log n)}$ SCGs. This reduces the worst-case complexity of the \LJB
algorithm to $2^{O(n \log n)}$. In the presence of this property, the massive
gap between Ramsey-based algorithms and rank-based algorithms vanishes, helping
to explain the surprising strength of the \LJB algorithm.

\begin{lem}\label{Rev_Det_SVW_Fast}
When operating on reverse-deterministic SCT problems, the \LJB algorithm has a
worst-case complexity of $2^{O(n \log n)}$. 
\end{lem}
\begin{proof}
A reverse-deterministic SCT problem contains only reverse-deterministic
SCGs. Observe that the composition of two
reverse-deterministic SCGs is itself reverse-deterministic. As there are only
$2^{O(n \log n)}$ possible reverse-deterministic SCGs, the closure computed in
the LJB algorithm cannot become larger than $2^{O(n \log n)}$. The LJB algorithm
checks each graph in the closure exactly once, and so has a time complexity of
$2^{O(n \log n)}$.
\end{proof}

It is not a coincidence that all SCT problems considered possess this property. 
As noted in \cite{BL07}, straightforward analysis of functional programs
generates only reverse-deterministic problems. In fact, every tool we examined is
only capable of producing reverse-deterministic SCT problems.
To illuminate the reason for this, imagine a
SCG $G : f \rightarrow g$ where $f$ has  two parameters, $x$ and $y$, and $g$
the single parameter $z$. If $G$ is not reverse deterministic, this implies both
$x$ and $y$ have arcs, labeled with either 0 or 1, to $z$.  This would mean that
$z$'s value is both \emph{always} smaller than or equal to $x$ and \emph{always}
smaller than or equal to $y$. 

The program in Algorithm \ref{Alg:GCD} can produce non-reverse-deterministic
size-change graphs, and serves to demonstrate the difficult analysis required to
do so\footnote{This example emerges from the Terminweb experiments by Mike
Codish, and was translated into a functional language by Amir Ben-Amram and Chin
Soon Lee. The authors are grateful to Amir Ben-Amram for bringing this
illustrative example to our attention.}.  Consider the SCG for the call on line
\ref{Alg:GCD:Call}. It is clear there should be a $0$-labeled arc from $X$ to
$X$. To reach this point, however, we must satisfy the inequality on line
\ref{Alg:GCD:Inequality}. Therefore we can also assert that $Y>X$, and include a
$1$-labeled arc from $Y$ to $X$. This is a kind of analysis is difficult to
make, and none of the size-change analyzers we examined were capable of
detecting this relation.

\LinesNumbered
\SetKwFunction{GCD}{gcd}
\begin{algorithm}[tp]
%\nofloatname
\DontPrintSemicolon
\caption{\FuncSty{gcd($X$,$Y$)}}
\label{Alg:GCD}
  \If{$Y>X$\label{Alg:GCD:Inequality}}
  {
     \GCD{$X$, $Y-X$}\label{Alg:GCD:Call}\;
  }
  \ElseIf{$X>Y$}
  {
     \GCD{$X-Y$,$Y$}\;
  }
  \lElse
  {
     \Return $X$
  }
\end{algorithm}

\subsection{Reverse Determinism and Rank-Based Containment}

Since the Ramsey-based approach benefited so strongly from reverse-determinism,
we examine the rank-based approach to see if it can similarly benefit.  As a
first step, we demonstrate that reverse-deterministic automata have a maximum
rank of 2, dramatically lowering the complexity of complementation to $2^{O(n)}$. We note, however,
that given a reverse-deterministic SCT problem $L$,i the automaton $\A_{Desc(L)}$ is \emph{not}
reverse-deterministic.  Thus a separate proof is provided to demonstrate that the rank of the
resulting automata is still bounded by 2. 

An automaton is \emph{reverse-deterministic} when no state has two incoming arcs
labeled with the same character. Formally, an automaton is reverse-deterministic
when, for each state $q$ and character $a$, there is at most one state $p$ such
that $q \in \rho(p,a)$. As a corollary to Lemma \ref{Rev_Det_SVW_Fast}, the
Ramsey-based complementation construction has a worst-case complexity of
$2^{O(n\log n)}$ for reverse deterministic automata With reverse-deterministic automata, we do not
have to worry about multiple paths to a state. As a consequence, a maximum rank of 2, rather than
$2n-2$, suffices to prove termination of every path, and the worst-case bound of the rank-based
construction improves to $2^{O(n)}$.

\begin{theorem}\label{Rev_Det_Rank_2}
Given a reverse-deterministic \buchi automaton $\B$ with $n$ states, there
exists an automaton $\B'$ with $2^{O(n)}$ states such that
$L(\B')=\overline{L(\B)}$.
\end{theorem}
\begin{proof}
In a run DAG $G_w$ of a reverse-deterministic automaton, all nodes
have only one predecessor. This implies the run DAG is a tree, and that the number of infinite paths
grows monotonically and at some point stabilizes. Call this point $k$.  If $G_w$ is rejecting, we
demonstrate that there is a point $j \geq k$ past which all accepting states are finite in $G_w$. 
Observe that each infinite path eventually stops visiting accepting states.  Let $j$ be the
last such point over all infinite paths, or $k$, whichever is greater. Past $j$, consider a branch
off this path containing an accepting state.  This branch cannot be a new infinite path, as the
number of infinite paths is stable.  This branch cannot lead to an existing infinite path, because
that would violate reverse determinism.  Therefore this path must be finite, and the accepting state
is finite.

Recall that $G_w(0)$ is $G_w$, $G_w(1)$ is $G_w(0)$ with all finite nodes removed, $G_w(2)$ is
$G_w(1)$ with all $F$-free nodes removed, and $G_w(3)$ is $G_w(2)$ with all finite nodes removed.
Because there are no infinite accepting nodes past $j$, $G_w(1)$ has no accepting nodes at all past
$j$. Thus every node past $j$ is $F$-free in $G_w(1)$, and $G_w(2)$ has no nodes past $j$. Thus
$G_w(3)$ is empty, and the DAG has a rank of at most $2$.  We conclude that the maximum rank of
rejecting run $DAG$ is 2, and the state space of the automaton in Definition \ref{KVDef} can be
restricted to level rankings with no ranking larger than 2.
\end{proof}

Unfortunately, neither the reduction of Definition \ref{LJB_Reduction} nor the
reduction of Definition \ref{Desc_Reduction} preserve reverse determinism,
which is to say that given a reverse-deterministic SCT problem, they do not
produce a reverse-deterministic \buchi containment problem.  However, we can
show that, given a reverse-deterministic SCT problem, the automata produced by
Definition \ref{Desc_Reduction} does have a maximum rank of 2. A similar claim
could be made about Definition \ref{LJB_Reduction} with minor adjustments.

\begin{figure}[tb]
\begin{center}
\includegraphics[width=0.4\linewidth]{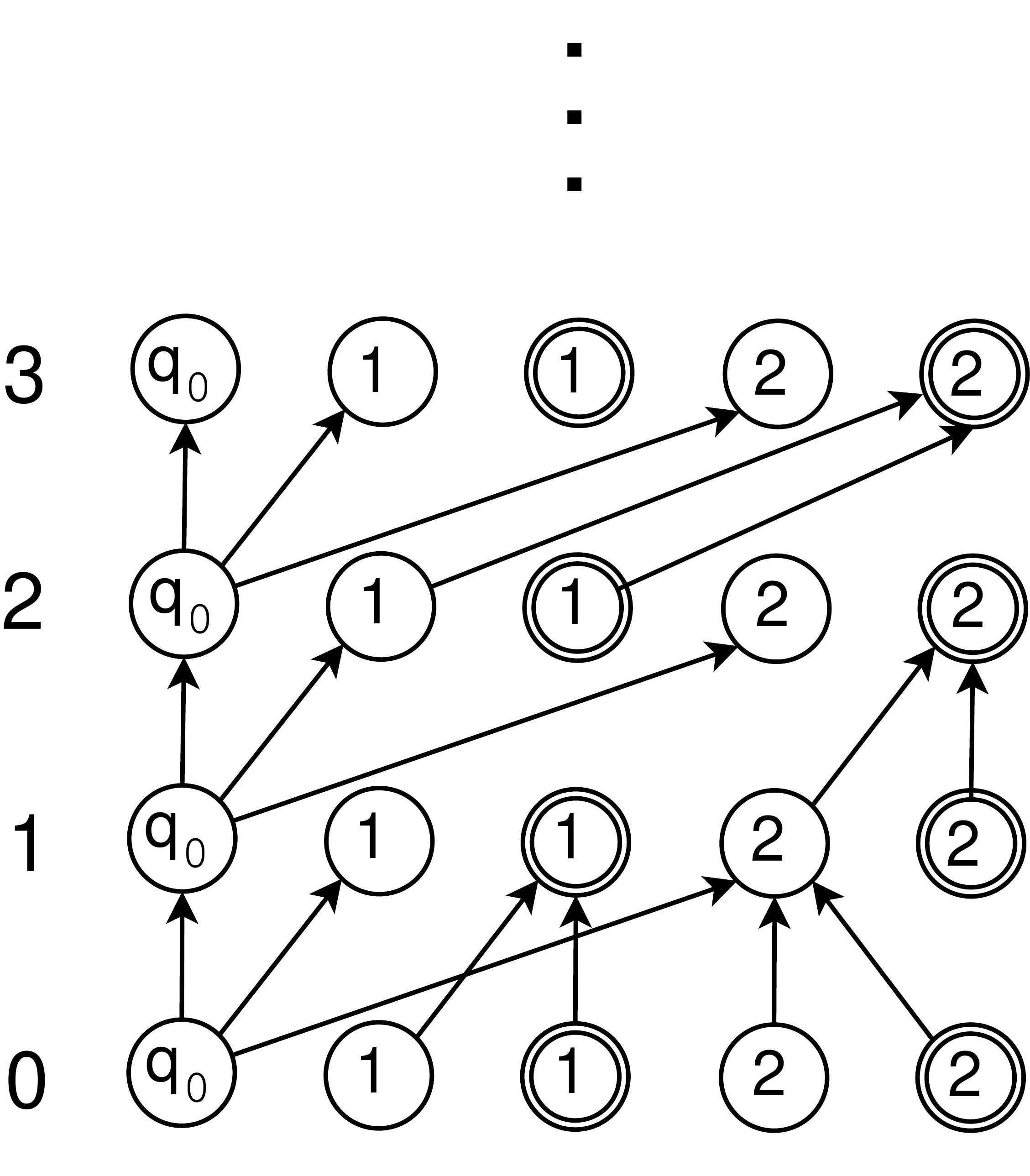}
\end{center}
\caption{An overapproximation of the run DAG for $\A'_{Desc(L)}$, where the maximum arity of
functions in $L$ $2$. For clarity, this figure includes unreachable states.  By definition, however,
the DAG has only nodes that can be reached from a node on the first level.  }\label{fig:Dag_Rank_2}
\end{figure}

Formally, we prove that for every reverse-deterministic SCT problem $L$, $\A'_{Desc(L)}$ has a
maximum rank of 2. Let $w$ be an infinite word $c_1c_2...$ not in $L(\A'_{Desc(L)})$, and $G_w$ the
rejecting run DAG of $\A'_{Desc(L)}$ on $w$.  There are two kinds of states in $\A'_{Desc(L)}$.
There is a waiting state, $q_0$, which always transitions to itself, and there are two states for
every variable position $h \in 1..n$, $\zug{h,0}$ and $\zug{h,1}$. Every state is an initial state.
Consider $G_w$, the run DAG of $\A'_{Desc(L)}$ on a word $c_0c_1...$.  Each character $c_i$
represents a function call from some function $f_i$ to another function $f_{i+1}$. At level $i$ of
the run DAG, the waiting state has outgoing edges to itself and the positions of 0-labeled
parameters of $f_{i+1}$. Each variable state only has outgoing edges to a 0 or 1-labeled position.
To get an idea of what the run DAG looks like, Figure \ref{fig:Dag_Rank_2} displays a supergraph of
the run DAG that includes all states at all levels, even if they are not reachable.

We now prove that the rejecting run DAG $G_w$ has a maximum rank of 2. To do so, we analyze the
structure of $G_w$ by first examining subgraphs, and then extending these observations to $G_w$. Let
$G_w'$ be the subgraph of the run DAG that omits the waiting state $q_0$ at every level of the run
DAG. Every path in $G_w'$ corresponds to a (possibly finite) thread in the call sequence
$c_1c_2\ldots$. 

\begin{lem}\label{Near_Tree}
For a level $i$, $f \in F$, and $x \in P(f)$, at most one of $(\zug{Ar(x),0},i)$ and
$(\zug{Ar(x),1},i)$ has incoming edges in $G_w'$.
\end{lem}
\begin{proof}
For $i=0$, this holds trivially. For $i>0$, take a pair of nodes
$(\zug{h,0},i)$ and $(\zug{h,1},i)$. The edges from level $i-1$ correspond to
transitions in $\A'_{Desc(L)}$ on some call $c$. As $c$ is a call to a single
function, we know there is a unique variable $x$ such that $h= Ar(x)$. Because
$L$ is reverse deterministic, we know that there is at most one edge in $\G_c$
leading to $x$. If there is no edge, then there are no edges entering
$(\zug{h,0},i)$ or $(\zug{h,1},i)$.

Otherwise there is exactly one edge in $\G_c$, $y \vararrow{r} x$, $r \in
\{0,1\}$.  In this case, the only nodes in level $G_w'$ with an edge to either
$(\zug{h,0},i)$ or $(\zug{h,1},i)$ are $(\zug{Ar(y),0},i-1)$ and
$(\zug{Ar(y),1},i-1)$.  By the transition function $\rho_D$, both of these
states transition only to $(\zug{h,r},i)$, and only $(\zug{h,r},i)$ has
incoming edges in $G_w'$.
\end{proof}

Now, define $G_w''$ to be the subgraph of $G_w'$ containing only nodes with an incoming edge in
$G_w'$. This removes nodes whose only incoming edge was from $q_0$.  While this excludes nodes that
begin threads, this cannot change the accepting or rejecting nature of a thread. 

\begin{lem}\label{Forest}\ 
\begin{enumerate}[\em(1)]
\item $G_w''$ is a forest.
\item Every infinite path in $G_w'$ appears in $G_w''$.
\item Every accepting node in $G_w$ is also in $G_w''$.
\end{enumerate}
\end{lem}
\begin{proof}\ \\
\standout{(1)}: Lemma \ref{Near_Tree} implies, for every $h$ and $i$, that only one of $(\zug{h,1},i)$ and
$(\zug{h,0},i)$ is in $G_w''$. Combined with the fact that $L$ is
reverse deterministic, every node in $G_w''$ can have at most one incoming
edge, and thus it is a forest.\\ 
\standout{(2)}: For every infinite path in $G_w'$, all nodes past the first have an
incoming edge from $G_w'$. Every node with an incoming edge from $G_w'$ is in
$G_w''$.  Thus for every infinite path in $G_w'$, a corresponding path, perhaps
without the first node, occurs in $G_w''$.\\
\standout{(3)}: Only nodes of the form $(\zug{h,1},i)$ are accepting. Let $(\zug{h,1},i)$
be a accepting node. By the definition of a run DAG, $(\zug{h,1},i)$ must be
reachable if it is in $G_w$. Thus there is an edge from another node in $G_w$
to $(\zug{h,1},i)$.  By the transition function , the waiting state $(q_0,i-1)$
only has an edge to $(\zug{h,0},i)$. Therefore the only nodes that have an edge
to $(\zug{h,1},i)$ are nodes of the form $(\zug{h',r},i-1)$. All nodes of this
form are in $G_w'$, and therefore $(\zug{h,1},i)$ has an incoming edge from
$G_w'$, and is in $G_w''$.
\end{proof}

We can now make observations about the rejecting run DAGs of $\A'_{Desc(L)}$ that mirror those made
about rejecting run DAGs of reverse deterministic automata.

\begin{lem}\label{Watershed}
There exists $j \in \N$ where for all $i > j$, $G_w''$ has no infinite
accepting nodes at level $i$.
\end{lem}
\begin{proof}
As $G_w''$ is a forest, the number of infinite nodes in level $i+1$ cannot
be smaller than the number of infinite nodes in level $i$.
Thus at some level $k$ the number of infinite nodes reaches a maximum.
Past level $k$, each infinite node has a unique infinite path through $G_w''$.
As $G_w$ is rejecting, every infinite path eventually stops visiting
accepting nodes at some level.
Let $j$ be the last such point over all infinite paths, or $k$, whichever
is greater.  
Past $j$, consider an accepting node $v$ that branch off an infinite path.
This branch cannot be part of the existing infinite path, as this path
has ceased visiting accepting nodes. Likewise, this branch cannot be part of a
new infinite path, as the number of infinite paths can not increase. Therefore
$v$ must be finite.
\end{proof}

\begin{lem}\label{Complete}
$G_w(3)$ is empty.
\end{lem}
\begin{proof}
Let $j$ be the level past which there are no infinite accepting nodes in
$G_w''$, as per Lemma \ref{Watershed}.  This precisely means that, past $j$,
every accepting node in $G_w''$ has a finite path.  As all accepting nodes in
$G_w$ are in $G_w''$, past $j$ every reachable accepting node in $G_w$ has a
finite path.  After level $j$, $G_w(1)$ contains only non-accepting nodes.  This
implies that $G_w(2)$ contains no nodes past $j$, and therefore that $G_w(3)$ is
empty.
\end{proof}

\begin{theorem}\label{Rev_Det_SCT_Rank_2}
Given a reverse-deterministic SCT problem $L$ with maximum arity $n$, there is
an automaton $\B'$ with at most $2^{O(n)}$ states such that
$L(\B')=\overline{L(\A'_{Desc(L)})}$.
\end{theorem}
\begin{proof}
By Lemma \ref{Complete}, we know that every rejecting run DAG of $\A'_{Desc(L)}$
has a maximum rank of 2. Therefore it suffices to restrict the rank in Definition
\ref{KVDef} to 2, replacing all occurrences of $2n-2$ in the definition with $2$.
The resulting automata is of size $O(2^{3n})$.
\end{proof}

\subsection{Experiments revisited}

In light of this discovery, we revisit the experiments and again compare rank
and Ramsey-based approaches on SCT problems. This time we tell Mh, the
rank-based solver, that the problems have a maximum rank of 2.  
Table \ref{Table:SCT_Completion_Revisited} compares the running time of Mh and
sct/scp on the five most difficult problems.  As before, time taken to reduce
SCT problems to automata containment problems was not counted.

\begin{table}[!htp]
%\vspace{0.04in}
\centering
\begin{tabular}{|l|c|c|}
\hline Problem & Mh (s) & sct/scp (s)\\
\hline ex04 & 0.01 & 1.39 \\
\hline ex05 & 0.13 & 227.7 \\
\hline ms &  0.1 & 0.02 \\
\hline gexgcd &  0.39 & 0.023 \\
\hline graphcolour2 &  0.044 & 0.014 \\
\hline
\end{tabular}
\vspace{0.05in}
\caption{SCT problem completion time times by tool, exploiting reverse-determinism.}\label{Table:SCT_Completion_Revisited}
%\vspace{0.10in}
\end{table}

While our problem space is small, the theoretical worst-case bounds of
Ramsey and rank-based approach appears to be reflected in the table. The
Ramsey-based sct/scp completes some problems more quickly, but in the worst
cases of ex04 and ex05, sct/scp performs significantly more slowly than Mh. It
is worth noting, however, that the benefits of reverse-determinism on
Ramsey-based approaches emerges automatically, while rank-based approaches must
explicitly test for this property in order to exploit it. 

\subsection{Monotonicity Constraints: Termination Problems Lacking
  Reverse-Deter\-minism}

Monotonicity constraints \cite{CLS05} are a generalization of size-change
graphs. While an SCG for a call from $f$ to $g$ is bipartite, with edges
only from variables of $f$ to variables of $g$, monotonicity constraints allow
edges between any two variables, even of the same function. In addition, while SCGs
only have edges representing \emph{less than} and \emph{less than or equal}
relations, monotonicity constraints allow edges representing \emph{equality}
relations. A collection of monotonicity constraints is called a monotonicity
constraint system (MCS). For a formal presentation, please see \cite{BenAm10}.

Deciding termination for MCS problems is more involved than for SCT problems,
but correctness similarly relies on Ramsey's Theorem \cite{CLS05}. One method
is to reduce a MCS to an SCT problem through {\em elaboration} \cite{BenAm10}. Unfortunately,
elaboration is an exponential reduction, and increases the size of the MCS.
Alternatively, it is possible to project an individual monotonicity constraint into an SCG in a
lossy fashion. To do so, simply remove all edges that are not from a variable of $f$ to a variable
of $g$, and replace equality edges with less-than-or-equal edges. By projecting every monotonicity
constraint in an MCS down to a SCG, we obtain a SCT problem.  Doing so, however, often removes
valuable information that can still be encoded in a size-change graph. To preserve this information,
new arcs that are logically implied by existing arcs can be added to the monotonicity constraint
before the constraint is projected to a SCG.  The simplest implied arcs are those derived from
equality edges: given two arcs $x \vararrow{a} y$ and $x \vararrow{=} x'$, add $x' \vararrow{a} y$
to the monotonicity constraint. Similarly, given an arc $y \vararrow{=} y'$, add $x \vararrow{a}
y'$. More complex implied arcs can be computed by similarly composing other arcs.

We obtained a corpus of 373 monotonicity constraint systems from
\cite{CGBFG11}. In each case, we produced three SCT problems from each MCS: one from directly projecting, one by computing arcs implied by
equality before projecting, and one by computing all implied arcs before
projecting. We again defined a problem to be interesting if either sct/scp or
Mh took more than 1 second to solve the problem. For every interesting problem,
there was no difference in result and no significant difference in running time
between the two types of implied arcs. Thus we consider only the third, most
complex, SCT problem generated from each MC problem, resulting in nine final
problems.

None of the interesting SCT problem produced in this fashion were reverse
deterministic. Given the complexity of monotonicity constraints, this is
perhaps unsurprising. Four of the resulting problems were non-terminating.
For these problems, the maximum rank can be computed. To do so, Mh is initially limited to a rank of
1, and the rank is increased until Mh can detect non-termination.  Table \ref{Table:MC_Completion}
displays the results for these problems. Despite the lack of reverse-determinism, none of these
problems proved difficult for sct/scp: consuming at most 0.4 seconds. However, several were
difficult for Mh, including one that took over eight minutes. In cases where we could bound the
rank, the running time for Mh often improved dramatically. While we again have only a sparse corpus
of interesting problems, these results serve to emphasize the importance of reverse determinism.
Perhaps more interestingly, they suggest that, even in cases where reverse determinism does not
hold, the Ramsey-based approach performs well.

\begin{table}[!htp]
%\vspace{0.04in}
\centering
\begin{tabular}{|l|c|c|c|}
\hline Problem & rank & Mh (s) & sct/scp (s) \\
\hline Test3 & N/A & 4.44 & 0.047 \\
\hline Test4 & N/A & 4.65 & 0.079 \\
\hline Test5 & N/A & 111.8 & 0.074 \\
\hline Test6 & N/A & 482.0 & 0.097 \\
\hline WorkingSignals & 13 & 1.32 (1.0) & 0.098 \\
\hline Gauss & 3 & 1.10 (0.08) & 0.146 \\
\hline PartitionList & 3 & 1.38 (0.22) & 0.081 \\
\hline Sudoku & 5 & 7.18 (2.42) & 0.405 \\
\hline
\end{tabular}
\vspace{0.05in}
\caption{MC problem, maximum rank for non-terminating problems, and completion times by tool. Times
for Mh in parenthesis are times when given the maximum rank, as if it were precomputed.}\label{Table:MC_Completion}
%\vspace{0.10in}
\end{table}

\section{Conclusion}\label{Sect:Conclusion}
In this paper we demonstrate that the Ramsey-based size-change
termination algorithm proposed by Lee, Jones, and Ben-Amram \cite{LJB01} is a
specialized realization of the 1987 Ramsey-based complementation construction
\cite{Buc62,SVW85}.  With this link established, we compare rank-based and
Ramsey-based tools on the domain of SCT problems. Initial experimentation
revealed a surprising competitiveness of the Ramsey-based tools, and led us to
further investigation. We discover that SCT problems are naturally
reverse-deterministic, reducing the complexity of the Ramsey-based approach. By
exploiting reverse determinism, we were able to demonstrate the superiority of
the rank-based approach.

Our initial test space of SCT problems was unfortunately small, with only five
interesting problems emerging. Despite the very sparse space of problem, they
still yielded two interesting observations. First, subsumption appears to be
critical to the performance of \buchi complementation tools using both rank and
Ramsey-based algorithms. It has already been established that rank-based tools
benefit strongly from the use of subsumption \cite{DR07}. Our results
demonstrate that Ramsey-based tools also benefit from subsumption, and in fact
experiments with removing subsumption from sct/scp seem to limit its
scalability. Second, by exploiting reverse determinism, we can dramatically
improve the performance of both rank and Ramsey-based approaches to containment
checking.

Reverse determinism, however, is not the whole story in comparing the rank and
Ramsey based approaches. On a separate corpus of problems derived from Monotonicity Constraints,
which are not reverse-deterministic, the Ramsey-based approach outperformed the rank-based
approach in every interesting case. It should be noted that, in addition to
reverse determinism, there are several ways to achieve a better bound on the
maximum rank than $2n-2$ \cite{FKV04, GKSV03}, even for problems that are not
known to be non-terminating. The rank-based approach might prove more competitive if such analyses
were applied before checking containment. None the less, it is clear that despite the theoretical
differences in complexity, we cannot discount the Ramsey-based approach. The competitive performance
of Ramsey-based solutions remains intriguing. 

In \cite{DR07,TV05}, a space of random automata universality problems is used to provide a diverse
problem domain. Unfortunately, it is far more complex to similarly generate a space of random SCT
problems. First, universality involves a single automaton: SCT problems check the containment of two
automata, with a corresponding increase in parameters. Worse, there is no reason to expect that one
random automaton will have any probability of containing another random automaton.  Sampling this
problem space is further complicated by the low transition density of reverse-deterministic
problems: in \cite{DR07,TV05} the most interesting problems had a transition density of 2.

On the theoretical side, we have extended the subsumption relation
present in sct/scp. Recent work has extended the subsumption relation to the
double-graph search of Algorithm \linebreak[4]\DGS, and others have improved the relation through the use of
simulation \cite{ACHMV10,FV10}. Doing so has enabled us to compared Ramsey and rank-based approaches
on the domain of random universality problems \cite{FV10}, with promising results.  Future work will
investigate how to generate an interesting space of random containment problems, addressing the
concerns raised above. 

The effects of reverse-determinism on the complementation of automata bear
further study.  Reverse-determinism is not an obscure property, it is known that
automata derived from LTL formula are often reverse-deterministic \cite{ES84b}.  As
noted above, both rank and Ramsey-based approaches improves exponentially when
operating on reverse-deterministic automata. Further, Ben-Amram and Lee have
defined SCP, a polynomial-time approximation algorithm for SCT \cite{BL07}. For a wide
subset of SCT problems with restricted in degrees, including the set used in
this paper, SCP is exact.  In terms of automata, this property is similar,
although perhaps not identical, to reverse-determinism.  The presence of an
exact polynomial algorithm for the SCT case suggests a interesting subset of
\buchi containment problems may be solvable in polynomial time. The first step
in this direction would be to determine what properties a containment problem
must have to be solved in this fashion.

\bibliography{ok,cav,sfogarty}

\begin{thebibliography}{10}

\bibitem{Daedalus}
Daedalus.
\newblock Available on:
  \url{http://www.di.ens.fr/~cousot/projects/DAEDALUS/index.shtml}.

\bibitem{ACHMV10}
P.A. Abdulla, Y.-F. Chen, L.~Hol\'{\i}k, R.~Mayr, and T.~Vojnar.
\newblock When simulation meets antichains.
\newblock In {\em Tools and Algorithms for the Construction and Analysis of
  Systems}, volume 6015 of {\em Lecture Notes in Computer Science}, pages
  158--174. Springer, 2010.

\bibitem{BenAm10}
A.M. Ben-Amram.
\newblock Size-change termination, monotonicity constraints and ranking
  functions.
\newblock {\em Logical Methods in Computer Science}, 6(3), 2010.

\bibitem{BL07}
A.M. Ben-Amram and C.S. Lee.
\newblock Program termination analysis in polynomial time.
\newblock {\em ACM Trans. Program. Lang. Syst.}, 29:5:1--5:37, January 2007.

\bibitem{Buc62}
J.R. B{\"u}chi.
\newblock On a decision method in restricted second order arithmetic.
\newblock In {\em Proc. Int. Congress on Logic, Method, and Philosophy of
  Science. 1960}, pages 1--12. Stanford University Press, 1962.

\bibitem{Cho74}
Y.~Choueka.
\newblock Theories of automata on $\omega$-tapes: A simplified approach.
\newblock {\em Journal of Computer and Systems Science}, 8:117--141, 1974.

\bibitem{CGBFG11}
M.~Codish, I.~Gonopolskiy, A.M. Ben-Amram, C.~Fuhs, and J.~Giesl.
\newblock {SAT}-based termination analysis using monotonicity constraints over
  the integers.
\newblock {\em Theory and Practice of Logic Programming, 26th Int'l. Conference
  on Logic Programming (ICLP'11) Special Issue}, 11:503--520, July 2011.

\bibitem{CLS05}
M.~Codish, V.~Lagoon, and P.J. Stuckey.
\newblock Testing for termination with monotonicity constraints.
\newblock In {\em Twenty First International Conference on Logic Programming},
  volume 3668 of {\em Lecture Notes in Computer Science}, pages 326--340.
  Springer-Verlag, October 2005.

\bibitem{DR07}
L.~Doyen and J.-F. Raskin.
\newblock Improved algorithms for the automata-based approach to
  model-checking.
\newblock In {\em Tools and Algorithms for the Construction and Analysis of
  Systems}, volume 4424 of {\em Lecture Notes in Computer Science}, pages
  451--465. Springer, 2007.

\bibitem{DR09}
L.~Doyen and J.-F. Raskin.
\newblock Antichains for the automata-based approach to model-checking.
\newblock {\em Logical Methods in Computer Science}, 5(1), 2009.

\bibitem{EL86}
E.A. Emerson and C.-L. Lei.
\newblock Efficient model checking in fragments of the propositional
  $\mu$-calculus.
\newblock In {\em Proc.\ 1st IEEE Symp. on Logic in Computer Science}, pages
  267--278, 1986.

\bibitem{ES84b}
E.A. Emerson and A.P. Sistla.
\newblock Deciding full branching time logics.
\newblock {\em Information and Control}, 61(3):175--201, 1984.

\bibitem{Fog08}
S.~Fogarty.
\newblock B{\"u}chi containment and size-change termination.
\newblock Master's thesis, Rice University, 2008.

\bibitem{FV10}
S.~Fogarty and M.Y. Vardi.
\newblock Efficient {B}{\"u}chi universality checking.
\newblock In {\em Tools and Algorithms for the Construction and Analysis of
  Systems}, volume 6015 of {\em Lecture Notes in Computer Science}, pages
  205--220. Springer, 2010.

\bibitem{Fred01}
C.C. Frederiksen.
\newblock A simple implementation of the size-change termination principle.
\newblock Available at:
  \url{ftp://ftp.diku.dk/diku/semantics/papers/D-442.ps.gz}, 2001.

\bibitem{FKV04}
E.~Friedgut, O.~Kupferman, and M.Y. Vardi.
\newblock B{\"u}chi complementation made tighter.
\newblock In {\em 2nd Int. Symp. on Automated Technology for Verification and
  Analysis}, volume 3299 of {\em Lecture Notes in Computer Science}, pages
  64--78. Springer, 2004.

\bibitem{Glen99}
Arne~J. Glenstrup.
\newblock Terminator {II}: Stopping partial evaluation of fully recursive
  programs.
\newblock Master's thesis, DIKU, University of Copenhagen, June 1999.

\bibitem{GKSV03}
S.~Gurumurthy, O.~Kupferman, F.~Somenzi, and M.Y. Vardi.
\newblock On complementing nondeterministic {B\"uchi} automata.
\newblock In {\em Proc. 12th Conf. on Correct Hardware Design and Verification
  Methods}, volume 2860 of {\em Lecture Notes in Computer Science}, pages
  96--110. Springer, 2003.

\bibitem{JB04}
N.D Jones and Nina Bohr.
\newblock Termination analysis of the untyped $\lambda$-calculus.
\newblock In {\em Rewriting Techniques and Applications. Proceedings}, volume
  3091 of {\em Lecture Notes in Computer Science}, pages 1--23. Springer, 2004.

\bibitem{Kla90}
N.~Klarlund.
\newblock {\em Progress Measures and finite arguments for infinite
  computations}.
\newblock PhD thesis, Cornell University, 1990.

\bibitem{KV97b}
O.~Kupferman and M.Y. Vardi.
\newblock Weak alternating automata are not that weak.
\newblock In {\em Proc. 5th Israeli Symp. on Theory of Computing and Systems},
  pages 147--158. {IEEE} Computer Society Press, 1997.

\bibitem{KV01}
O.~Kupferman and M.Y. Vardi.
\newblock Synthesizing distributed systems.
\newblock In {\em Proc.\ 16th IEEE Symp. on Logic in Computer Science}, pages
  389--398, 2001.

\bibitem{KV01c}
O.~Kupferman and M.Y. Vardi.
\newblock Weak alternating automata are not that weak.
\newblock {\em ACM Transactions on Computational Logic}, 2(2):408--429, 2001.

\bibitem{LJB01}
C.S. Lee, N.D. Jones, and A.M. Ben-Amram.
\newblock The size-change principle for program termination.
\newblock In {\em Proc.\ 28th ACM Symp. on Principles of Programming
  Languages}, pages 81--92, 2001.

\bibitem{Mic88}
M.~Michel.
\newblock Complementation is more difficult with automata on infinite words.
\newblock CNET, Paris, 1988.

\bibitem{Saf88}
S.~Safra.
\newblock On the complexity of $\omega$-automata.
\newblock In {\em Proc.\ 29th IEEE Symp. on Foundations of Computer Science},
  pages 319--327, 1988.

\bibitem{Sch09}
S.~Schewe.
\newblock B{\"u}chi complementation made tight.
\newblock In {\em 26th Int. Symp. on Theoretical Aspects of Computer Science},
  volume~3, pages 661--672. Schloss Dagstuhl, 2009.

\bibitem{SJ05}
D.~Sereni and N.D. Jones.
\newblock Termination analysis of higher-order functional programs.
\newblock In {\em 3rd Asian Symp. on Programming Languages and Systems}, volume
  3780 of {\em Lecture Notes in Computer Science}, pages 281--297. Springer,
  2005.

\bibitem{SVW85}
A.P. Sistla, M.Y. Vardi, and P.~Wolper.
\newblock The complementation problem for {B\"uchi} automata with applications
  to temporal logic.
\newblock In {\em Proc.\ 12th Int. Colloq. on Automata, Languages, and
  Programming}, volume 194, pages 465--474. Springer, 1985.

\bibitem{TV05}
D.~Tabakov and M.Y. Vardi.
\newblock Experimental evaluation of classical automata constructions.
\newblock In {\em Proc. 12th Int'l Conf. on Logic for Programming, Artificial
  Intelligence, and Reasoning}, Lecture Notes in Computer Science 3835, pages
  396--411. Springer, 2005.

\bibitem{Var07a}
M.Y. Vardi.
\newblock Automata-theoretic model checking revisited.
\newblock In {\em Proc. 8th Int. Conf. on Verification, Model Checking, and
  Abstract Interpretation}, volume 4349 of {\em Lecture Notes in Computer
  Science}, pages 137--150. Springer, 2007.

\bibitem{Wahl00}
David Wahlstedt.
\newblock Detecting termination using size-change in parameter values.
\newblock Master's thesis, G{\"o}teborgs Universitet, 2000.

\end{thebibliography}
\bibliographystyle{plain}

\end{document}